\documentclass[aps,prd,twocolumn,showpacs,superscriptaddress,floatfix]{revtex4-1}
\usepackage{graphicx,color}
\begin{document}

%\preprint{}

\title{Longitudinal double-spin asymmetry for inclusive jet and dijet production in $pp$ collisions at $\sqrt{s} = 510$ GeV}

\affiliation{Abilene Christian University, Abilene, Texas   79699}
\affiliation{AGH University of Science and Technology, FPACS, Cracow 30-059, Poland}
\affiliation{Alikhanov Institute for Theoretical and Experimental Physics, Moscow 117218, Russia}
\affiliation{Argonne National Laboratory, Argonne, Illinois 60439}
\affiliation{American Univerisity of Cairo, Cairo, Egypt}
\affiliation{Brookhaven National Laboratory, Upton, New York 11973}
\affiliation{University of California, Berkeley, California 94720}
\affiliation{University of California, Davis, California 95616}
\affiliation{University of California, Los Angeles, California 90095}
\affiliation{University of California, Riverside, California 92521}
\affiliation{Central China Normal University, Wuhan, Hubei 430079 }
\affiliation{University of Illinois at Chicago, Chicago, Illinois 60607}
\affiliation{Creighton University, Omaha, Nebraska 68178}
\affiliation{Czech Technical University in Prague, FNSPE, Prague 115 19, Czech Republic}
\affiliation{Technische Universit\"at Darmstadt, Darmstadt 64289, Germany}
\affiliation{E\"otv\"os Lor\'and University, Budapest, Hungary H-1117}
\affiliation{Frankfurt Institute for Advanced Studies FIAS, Frankfurt 60438, Germany}
\affiliation{Fudan University, Shanghai, 200433 }
\affiliation{University of Heidelberg, Heidelberg 69120, Germany }
\affiliation{University of Houston, Houston, Texas 77204}
\affiliation{Huzhou University, Huzhou, Zhejiang  313000}
\affiliation{Indian Institute of Science Education and Research (IISER), Berhampur 760010 , India}
\affiliation{Indian Institute of Science Education and Research, Tirupati 517507, India}
\affiliation{Indian Institute Technology, Patna, Bihar, India}
\affiliation{Indiana University, Bloomington, Indiana 47408}
\affiliation{Institute of Physics, Bhubaneswar 751005, India}
\affiliation{University of Jammu, Jammu 180001, India}
\affiliation{Joint Institute for Nuclear Research, Dubna 141 980, Russia}
\affiliation{Kent State University, Kent, Ohio 44242}
\affiliation{University of Kentucky, Lexington, Kentucky 40506-0055}
\affiliation{Lawrence Berkeley National Laboratory, Berkeley, California 94720}
\affiliation{Lehigh University, Bethlehem, Pennsylvania 18015}
\affiliation{Max-Planck-Institut f\"ur Physik, Munich 80805, Germany}
\affiliation{Michigan State University, East Lansing, Michigan 48824}
\affiliation{National Research Nuclear University MEPhI, Moscow 115409, Russia}
\affiliation{National Institute of Science Education and Research, HBNI, Jatni 752050, India}
\affiliation{National Cheng Kung University, Tainan 70101 }
\affiliation{Nuclear Physics Institute of the CAS, Rez 250 68, Czech Republic}
\affiliation{Ohio State University, Columbus, Ohio 43210}
\affiliation{Panjab University, Chandigarh 160014, India}
\affiliation{Pennsylvania State University, University Park, Pennsylvania 16802}
\affiliation{NRC "Kurchatov Institute", Institute of High Energy Physics, Protvino 142281, Russia}
\affiliation{Purdue University, West Lafayette, Indiana 47907}
\affiliation{Pusan National University, Pusan 46241, Korea}
\affiliation{Rice University, Houston, Texas 77251}
\affiliation{Rutgers University, Piscataway, New Jersey 08854}
\affiliation{Universidade de S\~ao Paulo, S\~ao Paulo, Brazil 05314-970}
\affiliation{University of Science and Technology of China, Hefei, Anhui 230026}
\affiliation{Shandong University, Qingdao, Shandong 266237}
\affiliation{Shanghai Institute of Applied Physics, Chinese Academy of Sciences, Shanghai 201800}
\affiliation{Southern Connecticut State University, New Haven, Connecticut 06515}
\affiliation{State University of New York, Stony Brook, New York 11794}
\affiliation{Temple University, Philadelphia, Pennsylvania 19122}
\affiliation{Texas A\&M University, College Station, Texas 77843}
\affiliation{University of Texas, Austin, Texas 78712}
\affiliation{Tsinghua University, Beijing 100084}
\affiliation{University of Tsukuba, Tsukuba, Ibaraki 305-8571, Japan}
\affiliation{United States Naval Academy, Annapolis, Maryland 21402}
\affiliation{Valparaiso University, Valparaiso, Indiana 46383}
\affiliation{Variable Energy Cyclotron Centre, Kolkata 700064, India}
\affiliation{Warsaw University of Technology, Warsaw 00-661, Poland}
\affiliation{Wayne State University, Detroit, Michigan 48201}
\affiliation{Yale University, New Haven, Connecticut 06520}

\author{J.~Adam}\affiliation{Brookhaven National Laboratory, Upton, New York 11973}
\author{L.~Adamczyk}\affiliation{AGH University of Science and Technology, FPACS, Cracow 30-059, Poland}
\author{J.~R.~Adams}\affiliation{Ohio State University, Columbus, Ohio 43210}
\author{J.~K.~Adkins}\affiliation{University of Kentucky, Lexington, Kentucky 40506-0055}
\author{G.~Agakishiev}\affiliation{Joint Institute for Nuclear Research, Dubna 141 980, Russia}
\author{M.~M.~Aggarwal}\affiliation{Panjab University, Chandigarh 160014, India}
\author{Z.~Ahammed}\affiliation{Variable Energy Cyclotron Centre, Kolkata 700064, India}
\author{I.~Alekseev}\affiliation{Alikhanov Institute for Theoretical and Experimental Physics, Moscow 117218, Russia}\affiliation{National Research Nuclear University MEPhI, Moscow 115409, Russia}
\author{D.~M.~Anderson}\affiliation{Texas A\&M University, College Station, Texas 77843}
\author{R.~Aoyama}\affiliation{University of Tsukuba, Tsukuba, Ibaraki 305-8571, Japan}
% ...

\collaboration{STAR Collaboration}\noaffiliation

\date{\today}

\begin{abstract}
\noindent We report the first measurement of the inclusive jet and the dijet longitudinal double-spin asymmetries, $A_{LL}$, at midrapidity in polarized $pp$ collisions at a center-of-mass energy $\sqrt{s} = 510$ GeV. The inclusive jet $A_{LL}$ measurement is sensitive to the gluon helicity distribution down to a gluon momentum fraction of $x\approx 0.015$, while the dijet measurements, separated into four jet-pair topologies, provide constraints on the $x$ dependence of the gluon polarization. Both results are consistent with previous measurements made at $\sqrt{s}= 200$ GeV in the overlapping kinematic region, $x > 0.05$, and show good agreement with predictions from recent next-to-leading order global analyses. 
\end{abstract}

%\keywords{}

\maketitle

\section{Introduction}
The proton consists of quarks and antiquarks, bound by gluons. The gluons provide about half of the momentum of the proton (see for example \cite{Dulat:2015mca}), and their interactions provide most of the mass \cite{Yang:2018nqn,Ji:1994av}.  Nonetheless, we know very little about the role that gluons play in determining the fundamental proton quantum numbers, such as its spin.

The spin program at the Relativistic Heavy Ion Collider (RHIC) has made significant progress toward addressing the question of how much, if at all, gluon spins contribute to the spin of the proton.  The STAR and PHENIX collaborations have performed a sequence of measurements of the longitudinal double-spin asymmetry, $A_{LL}$, for inclusive jet \cite{run3run4res2006,run6aLL2008,run6aLL2012,run9aLL2015} and pion \cite{PHENIXrun5,PHENIXrun6,PHENIXpp62,STARpi0,PHENIXrun9} production.  The results have been incorporated, along with inclusive and semi-inclusive lepton-proton scattering data, into the recent DSSV14 \cite{DSSV14} and NNPDFpol1.1 \cite{NNPDFpol} next-to-leading order (NLO) perturbative QCD global analyses. These extractions of the helicity parton distribution functions (PDFs) indicate that, at momentum transfer scale of $Q^2=10$~(GeV/$c$)$^2$ and for momentum fractions $x > 0.05$ that are sampled by the included RHIC data, gluon spins contribute approximately $40\%$ of the total proton spin.

\begin{figure}[tbh]
\includegraphics[width=\columnwidth]{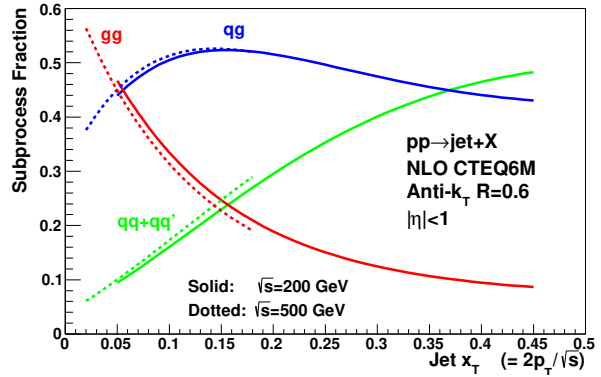}
\caption{\label{fig:subprocess}Fractions of the next-to-leading-order cross section \cite{cteq6l2002,nlojet2012} for inclusive jet production arising from quark-quark, quark-gluon, and gluon-gluon scattering in $pp$ collisions at $\sqrt{s} = 200$ and 500~GeV, as a function of $x_T = 2 p_T/\sqrt{s}$.}
\end{figure}

RHIC data provide direct, leading-order sensitivity to gluon polarization because hard scattering processes at RHIC energies are dominated by gluon-gluon and quark-gluon scattering, as shown in Fig.\@ \ref{fig:subprocess}.  In contrast, polarized lepton scattering data constrain the gluon polarization indirectly, via $Q^2$ evolution effects.  There have been two recent global analyses \cite{BS15,JAM15} that only included lepton scattering data in their fits.  These fits also find substantial gluon polarization in the region $x > 0.05$, albeit with larger uncertainties than those of \cite{DSSV14,NNPDFpol}. Recently, the first lattice QCD calculation of the full first moment of the gluon helicity distribution $\Delta g(x,Q^2)$ has been calculated to be $\Delta G(Q^2)= \int_0^1\Delta g(x,Q^2)dx$ = 0.251~$\pm$~0.047\,(stat.)~$\pm$~0.016\,(syst.) at $Q^2$=10~(GeV/$c$)$^2$~\cite{LatticeDG} .  In addition, the small-$x$ asymptotic behavior of $\Delta g(x)$ has been derived in the large-$N_c$ limit \cite{smallxglue}, although the $x$ range where the asymptotic limit is applicable is not yet clear.

While the DSSV14 and NNPDFpol1.1 analyses are in good agreement for the kinematic region $x>0.05$ where the included data from RHIC on inclusive jet and neutral pion production at $\sqrt{s} = 200$~GeV are most sensitive, the extrapolations over smaller $x$ and their associated errors are markedly different.  For example, at $x=10^{-3}$, the quoted gluon polarization uncertainty in NNPDFpol1.1 is twice as large as that for DSSV14. These extrapolations are needed to determine the full first moment of the gluon helicity distribution. Complementary measurements are thus required both to extend the sensitivity to smaller $x$ and better to resolve the $x$ dependence of $\Delta g(x,Q^2)$.

%  Not surprisingly, these same limitations result in an order of magnitude larger total uncertainty on the extracted $\Delta{G}(Q^2)$ compared to the uncertainty on $\Delta{\Sigma}(Q^2)$, the first moment of the sum of the quark and anti-quark helicity distributions. 

The inclusive jet and the dijet longitudinal double-spin asymmetries presented in this paper will help address both issues. The data for these measurements were collected from $\sqrt{s}=510$ GeV polarized $pp$ collisions during the 2012 RHIC running period. For a given jet transverse momentum, $p_T$, and pseudorapidity, $\eta$, the increased center-of-mass energy extends the sensitivity of the inclusive jet channel to lower $x$ partons ($x\simeq x_T e^{\pm\eta}$, where $x_T = 2 p_T/\sqrt{s}$). While the inclusive jet channel provides the strongest statistical power, dijets permit extraction of the momentum fractions, $x_1=(p_{T,3}e^{+\eta_3}+p_{T,4}e^{+\eta_4})/\sqrt{s}$ and $x_2=(p_{T,3}e^{-\eta_3}+p_{T,4}e^{-\eta_4})/\sqrt{s}$, of the partons participating in the hard scattering at the Born level, with higher-order corrections that are known and have been shown to be small \cite{DdF:hadronjet}.  Note that, throughout this paper, the kinematics of the initial partons and final jets are denoted by subscripts 1,2 and 3,4, respectively.  The $\sqrt{s}=510$ GeV dijet asymmetries here are separated into four pseudorapidity topologies that facilitate the extraction of $x$-dependent constraints as a function of the dijet invariant mass $M_{34} = \sqrt{sx_1x_2}$.  Together, these inclusive jet and dijet results will provide important new constraints on the magnitude and shape of the gluon polarization over the range $0.015 < x < 0.2$.  

A number of other measurements sensitive to gluon polarization have been released since the DSSV14 and NNPDFpol1.1 global analyses.  STAR has published the first two measurements of dijet $A_{LL}$, based on $pp$ collision data at $\sqrt{s}=200$~GeV.  One measurement considers asymmetries for dijets at midrapidity \cite{run9dijet200}, while the second considers cases where at least one jet falls within $0.8 < \eta < 1.8$ \cite{STAREEMCdijet}.  Very recently, an update of the DSSV14 fit has been performed that includes these two dijet measurements by reweighting \cite{DSSV:2019mc}.  The updated fit finds that inclusion of the STAR $\sqrt{s}=200$ GeV dijet results leads to a small increase in the size of $\Delta g(x)$ in the region $0.05<x<0.2$, together with a sizable reduction in the width of the uncertainty band, with the latter most notable in the region $x \gtrsim 0.2$.
%These will provide complementary constraints on the shape of $\Delta g(x,Q^2)$.  The latter measurement also provides sensitivity to gluon polarization down to $x \approx 0.01$.  

Asymmetries have also been measured for inclusive $\pi^0$ production in 510 GeV $pp$ collisions at $|\eta| < 0.35$ by PHENIX \cite{PHENIXpp510} and at $2.65 < \eta < 3.9$ by STAR \cite{STARFMSpi0}. These $\pi^0$ asymmetries provide sensitivity to gluon polarization down to $x \approx 0.01$ and $x \approx 0.001$, respectively.

The remainder of this paper is organized as follows. Section \ref{sect:Experiment} briefly describes the components of the RHIC complex and the Solenoidal Tracker at RHIC (STAR) detector that are relevant to this measurement. Section \ref{sect:Jetreco} discusses jet reconstruction, including an underlying event and background subtraction technique. Section \ref{sect:Embed} reviews the simulation sample that is used to correct the data for detector effects such as acceptance and resolution and to estimate systematic uncertainty contributions. Section \ref{sect:Asymmetries} discusses the determination of $A_{LL}$ and contributions to the systematic uncertainty. Section \ref{sect:Results} presents the results, along with comparisons to theoretical predictions. Section \ref{sect:conclusions} provides a brief conclusion.

\section{Experiment and Data}\label{sect:Experiment}

\subsection{The STAR detector at RHIC}
The RHIC complex has accelerated and collided beams of polarized protons at center-of-mass energies ranging from $62-510$~GeV \cite{rhicproj,rhicdesign,rhicpolarize}. During the 2012 running period, each beam was typically filled with 111 bunches of vertically polarized protons. Rotator magnets placed on either side of the STAR interaction region were used to rotate the proton spin orientation from vertical to longitudinal. To minimize systematic effects due to bunch-to-bunch variations, the helicity state assigned to a pair of colliding bunches (++, +$-$, $-$+, $--$) was varied through the 2012 running period. This design also allowed for the flipping of beam spin orientation at the same rate as the colliding bunches, on the order of once every 100 ns. For a given storage period, or fill, the polarization of each beam was measured several times using Coulomb-nuclear interference (CNI) proton-carbon polarimeters \cite{pCpol}. The CNI polarimeters were calibrated with a polarized atomic hydrogen gas-jet target \cite{hjetpol}. The luminosity-weighted polarizations for the two beams, which are referred to as ``blue'' and ``yellow'', were 54\% and 55\%. 

STAR is a large acceptance, multi-purpose detector located at the RHIC 6 o'clock interaction region \cite{rhicstar}. The detector components used in this analysis are the Time Projection Chamber (TPC), the Barrel (BEMC) and Endcap (EEMC) Electromagnetic Calorimeters (collectively, EMCs), the Vertex Position Detector (VPD), and the Zero Degree Calorimeters (ZDC). The TPC measures the momentum of charged particles scattered within $|\eta| \lesssim 1.3$ and $0 <\phi< 2\pi$ \cite{startpc}.  The EMCs measure the energy of photons, electrons, and positrons, and provide the triggering. The BEMC \cite{starbemc} spans the region $-1 < \eta < 1$, and the EEMC \cite{stareemc} spans $1.1 < \eta < 2$, both with full azimuth. The VPD and ZDC are pairs of far-forward, fast-response detectors east and west of the interaction region.  The VPD samples the region $4.2 < |\eta| < 5.2$ \cite{starvpd}, while the ZDCs cover $|\eta| > 6.6$ \cite{starzdc}. Hit information from the VPD and ZDC detectors is used to extract the relative luminosities of the colliding bunches associated with a given helicity state. Azimuthal segmentation in the ZDC also allows it to serve as a local polarimeter to verify that the rotator magnets are set and functioning properly.

\subsection{Triggers and event selection}
STAR sampled 82~pb$^{-1}$ of longitudinally polarized $pp$ collisions at $\sqrt{s}=510$~GeV during the 2012 running period. The inclusive jet and the dijet analyses both utilize jet patch (JP) triggers which are constructed by applying thresholds to the total transverse energy ($E_T$) detected within $\Delta\eta\times\Delta\phi= 1\times 1$ regions in the EMCs.  There are a total of 30 jet patches, with five patches that overlap in $\eta$ for each of six non-overlapping regions in $\phi$. An event satisfied the JP0, JP1 or JP2 trigger if the $E_T$ of at least one of the jet patches exceeded 5.4, 7.3 or 14.4 GeV, respectively.  All JP2-triggered events were recorded while the JP0 and JP1 triggers were prescaled to fit within the available data-acquisition bandwidth. During off-line processing, events are also required to pass a software trigger simulator that incorporates time-dependent pedestal variations and detector efficiencies. 

Candidate collision vertices are reconstructed from TPC tracks and hits in the EMCs and then ranked based on the number of in-time tracks and their transverse momenta. To ensure reasonable detector acceptance and minimize pileup events, only the highest quality vertex in each event is selected, and the position along the beam line, $z_\mathrm{vertex}$, is required to fall within $\pm 90$~cm of the center of the STAR detector.

The vertex reconstruction efficiency drops rapidly at the highest instantaneous luminosities achieved during the 2012 running period. The highest luminosity data are excluded from the inclusive jet analysis in order to minimize errors due to associating jets with the wrong vertex, as discussed in Sect.\@ \ref{subsubsect:TrigBias}. In contrast, these data are included in the dijet analysis, as the higher observed track multiplicity makes it much less likely to assign the wrong vertex to a dijet event.

\section{Jet reconstruction}\label{sect:Jetreco}

 A jet is a cluster of particles that originates from fragmentation and hadronization of an energetic final-state parton in a hard scattering process. They are abundantly produced at RHIC in 2$\rightarrow$2 QCD processes where an initial-state parton is freed from a polarized proton beam. At STAR, jets are reconstructed from the charged tracks measured by the TPC and the energy deposits in the EMCs. The STAR jet finding algorithms have evolved in step with advances of jet finding techniques in the community and with the increasingly complex experimental conditions that accompany higher center-of-mass energy collisions and luminosities. Early STAR $pp$ analyses \cite{run3run4res2006,run6aLL2008,run6aLL2012} implemented the mid-point cone algorithm \cite{conesalgo}.  The cone radius varied from $R=0.4-0.7$ as the EMCs acceptance was gradually expanded. The 2009 $\sqrt{s} = 200$~GeV inclusive jet analysis \cite{run9aLL2015} represented the first STAR results obtained with the anti-$k_T$ algorithm \cite{antikt}, a change that significantly reduces the sensitivity to soft background and pileup effects. The inclusive jet and dijet analyses presented in this paper also use the anti-$k_T$ algorithm, as implemented in \textsc{FastJet} version 3.0.6 \cite{fastjet}, but with a smaller jet resolution parameter, $R = 0.5$ \textit{vs}.\@ $R = 0.6$ at $\sqrt{s}=200$~GeV, to reduce the increased contributions from soft background at $\sqrt{s}=510$~GeV.

\subsection{Inputs to the jet finder}

The TPC tracks included in the jet finding algorithm are required to have at least 12 fit points out of a possible 45 to provide good momentum resolution. To remove split tracks, the number of hit points must be greater than 51\% of the maximum possible number when the track geometry and active electronic channels are considered.  In addition, tracks must have transverse momenta $p_T > 0.2$~GeV/$c$ and be associated with the selected vertex for the event within a $p_T$-dependent distance of closest approach (DCA).  The DCA is required to be less than 2~cm for $p_T < 0.5$~GeV/$c$ and less than 1~cm for $p_T > 1.5$~GeV/$c$; the DCA requirement is linearly interpolated between these two limits. The four-momenta of the charged tracks are constructed by equating the rest mass of each track to the pion mass. The EMC tower hits included in the jet finding algorithm are required to have a signal well above pedestal and an $E_T > 0.2$~GeV. The 4-momentum of an EMC hit is constructed by setting the rest mass to zero, as if all the energy deposited was due to photons originating from the vertex.

For tracks pointing to an EMC hit, the track $p_T$ (multiplied by $c$ to account for units) is subtracted from the tower $E_T$.  If the difference is less than zero, the tower is discarded from the jet finding. This procedure, which is referred to as ``$p_T$ subtraction'', avoids double counting from electrons and positrons that are fully reconstructed by both the TPC and EMCs.  In contrast, on average charged hadrons deposit only $\simeq$\,30\% of their energy in the EMC material.  Therefore, $p_T$ subtraction results in an over-subtraction in the rare case where a photon strikes the same tower as a charged hadron. However, by suppressing the sensitivity to the large event-to-event fluctuations in the charged hadron energy deposition, it significantly improves the resolution of the reconstructed jet energy \cite{run9aLL2015}.

\subsection{Underlying event subtraction}

The underlying event (UE) is composed of low-$p_T$ particles originating from multiple parton interactions and soft interactions between the scattered partons and proton remnants. The underlying event at RHIC energies is expected to be isotropic and approximately independent of the scale of the hard interaction~\cite{CdF:UE}. As a result, distortions to the energy scale are largest for low-$p_T$ jets. The UE has also been assumed to be spin independent~\cite{run9aLL2015}, but that assumption has not been verified experimentally before this work.

A technique, adapted from the ALICE experiment~\cite{alicecone}, is applied for each jet in the analysis to correct the underlying event contribution to the reconstructed jet $p_T$ and dijet invariant mass $M_{inv}$. The algorithm, called the off-axis method, scans the same list of TPC tracks and EMC hits that was input to the jet finder and selects those located in two off-axis cones, with radius $R = 0.5$ (chosen to match the anti-$k_T$ resolution parameter), centered at the same $\eta$ as the jet but $\pm \pi/2$ away in $\phi$.

For the inclusive jet analysis, the average transverse momentum density per unit area deposited inside the two cones, $\hat{\rho}$, is computed and the correction $dp_T = \hat{\rho} A_{jet}$ is applied to the jet $p_T$. $A_{jet}$ is the jet area and is given by the anti-$k_T$ algorithm using the ghost particle method~\cite{antikt}.

In the dijet analysis, the 4-momentum is calculated for the collection of particles in each off-axis cone, summed and then rotated by $\pm\pi/2$ back to the position of the jet. After rotation, the off-axis cone 4-momenta are averaged, scaled to the area of the jet $A_{jet}$, and subtracted from the initial jet 4-momentum. The underlying event correction is calculated and applied on a jet-by-jet basis for both the dijet and the inclusive jet analyses.

This technique recognizes that the STAR detector has excellent four-fold symmetry in azimuth, but the efficiency is not as uniform in pseudorapidity.  For example, there is a small gap in the EMC coverage between the BEMC and EEMC. Requiring the off-axis cones to be centered at the same $\eta$ as the jet and sum over similar areas ensures the $\eta$ dependence of the underlying event and other background contributions are sampled correctly, and facilitates the jet-by-jet correction.  It is important to note that, in addition to accounting for the UE, this procedure also corrects for pileup effects arising, for example, from beam-beam and beam-gas collisions other than the $pp$ collision of interest.

\subsection{Dijet and inclusive jet event selection}

Jets are selected for further analysis if the jet axis lies within $|\eta| < 0.9$ and the $p_T > 6$~GeV/$c$. To minimize jet energy corrections near the detector acceptance limits, an additional $z_\mathrm{vertex}$-dependent $\eta$ cut ensures each jet thrust axis projects well within the BEMC. The remaining cuts are tuned specifically for the inclusive jet or dijet analysis and are detailed in the following sections.

\subsubsection{Inclusive jet cuts}\label{sect:incl_jet_cuts}
The jets are divided into three mutually exclusive groups depending on the highest jet patch trigger a specific jet can satisfy: JP0, JP1, or JP2. For example, a jet that deposited enough energy in the EMC to satisfy the JP1 trigger requirement, but fired only the JP0 trigger because of the JP1 prescale, is nonetheless categorized as JP1 during the analysis. In addition, each jet is required to point toward a jet patch that could trigger the event, including the constraints from prescales. The minimum reconstructed jet $p_T$ values for the three trigger categories of 6.0, 8.2 and 15.3~GeV/$c$, respectively, are set at $p_T$ bin boundaries that are somewhat higher than the corresponding JP hardware trigger thresholds to reduce reconstruction bias near the trigger thresholds.

The summed transverse momenta of the charged tracks within a jet is required to be greater than 0.5~GeV/$c$, and the fraction of jet energy detected in the EMCs, $R_{EM} = E_{EM}/(E_{EM}+E_{track})$, is required to be less than 0.94.  These constraints suppress non-collision backgrounds such as cosmic events and beam backgrounds, which do not point back to the event vertex, and reduce the probability that the wrong vertex is assigned to the jet. The track momentum resolution degrades for $p_T > 30$~GeV/$c$, so jets with such tracks are rejected.

Approximately 5\% of the events in the inclusive jet analysis contain two jets, both of which satisfy all the cuts.  In these cases, both jets are considered.  Fewer than 0.05$\%$ of events have three or more jets that satisfy all the cuts. For these cases, only the two jets with the highest $p_T$ are considered.

\subsubsection{Dijet cuts}\label{sect:dijet_cuts}

Only events with two or more jets are considered for the dijet sample. From these events, the jets with the highest $p_T$ are selected as the candidate dijet pair. At least one jet of the pair is required to point to a jet patch that satisfies the JP0, JP1 or JP2 trigger and to pass an associated threshold of 6.0, 8.0 or 15.0~GeV/$c$, respectively. Both jets must have $R_{EM} < 0.95$.  The latter constraint can be less stringent than used for inclusive jets because dijet events are less susceptible to backgrounds.  The events are separated into three mutually exclusive groups (JP0, JP1, and JP2) using the same algorithm as for inclusive jets. 

A dijet opening-angle cut, $\Delta \phi > 120^{\circ}$, is designed to remove the cases where one member of the dijet pair is the result of a hard gluon emission.  An additional dijet opening-angle cut, $|\Delta \eta| < 1.6$, removes the kinematic region where both jets fall near the detector acceptance limits.  A $p_T$-matching condition is applied that requires the ratio of the leading and away-side jet transverse momenta, $p^{leading}_T/p^{away}_T < (6-0.08p^{max}_T)$, where $p^{max}_T$ is the transverse momentum of the highest $p_T$ track in either jet. This empirical cut was tuned on simulation and motivated by the need to remove fake jets that are composed nearly entirely of a single, poorly reconstructed TPC track. Finally, an asymmetric $p_T$ cut requires one of the jets to have a $p_T > 8$ GeV/$c$ and the other $p_T > 6$ GeV/$c$. The latter condition is motivated by theoretical considerations \cite{Asymdijet}. 

\begin{table}
\caption{The four dijet topology bins A-D.  In all cases, the dijet pair is also required to satisfy $\Delta \phi > 120^0$ and $|\Delta \eta| < 1.6$.}
\label{table:topological}
\begin{tabular}{ c @{~}|@{~~} c @{~~}|@{~} c }
  \hline\hline	
  Bin & $\eta_3$ and $\eta_4$ Regions & Physics Description \\
  \hline
  A & $0.3 < |\eta_{3,4}| < 0.9$; $\eta_3\cdot\eta_4 > 0$ & Forward-Forward \\
  B & $|\eta_{3,4}|<0.3$; $0.3 < |\eta_{4,3}| < 0.9$ & Forward-Central \\
  C & $|\eta_{3,4}|<0.3$ & Central-Central\\
  D & $0.3 < |\eta_{3,4}| < 0.9$; $\eta_3\cdot\eta_4 < 0$ & Forward-Backward \\
  \hline\hline  
\end{tabular}
\end{table}

Identified dijets are sorted into four topological bins (A-D) based on the pseudorapidities of the individual jets in the dijet pair. Three pseudorapidity regions are defined as follows: forward spanning $0.3<\eta<0.9$, central spanning $-0.3<\eta<0.3$, and backward spanning $-0.9<\eta<-0.3$. These three regions permit the construction of four unique topological bins, described in Table \ref{table:topological}. The dijets in bins A and B reflect the most asymmetric collisions in terms of partonic $x_1$ and $x_2$ and, therefore, sample the highest and lowest $x$ values. The dijets in bins C and D originate from more symmetric partonic collisions, and largely access intermediate $x$ values. Bins A and C sample collisions with $|\cos\theta^*|$ near zero, while bins B and D sample regions of larger $|\cos\theta^*|$. The factor $\cos\theta^*$, where $\theta^*$ is the scattering angle in the partonic center-of-mass frame, enters directly into the calculation of the partonic asymmetry, $\hat{a}_{LL}$.  For the present case, $\hat{a}_{LL}$ is larger when  $|\cos\theta^*|$ is smaller.

\section{Embedded Simulation}\label{sect:Embed}

A Monte Carlo simulation is used to determine corrections to the measured jet quantities and estimate contributions to the total systematic uncertainty. Simulated $pp$ events generated by \textsc{Pythia} \cite{pythia2006}  are passed through a detailed \textsc{Geant3} \cite{geant3} simulation of the STAR detector utilizing a geometry setup matching the 2012 detector configuration. The simulated detector responses are then embedded into zero-bias events, which were recorded without any trigger requirement at random times during the running period. In this way, the simulated events contain the same pile-up and beam backgrounds as the real data. After embedding, the simulated EMC tower ADCs are analyzed by the trigger simulator in order to identify those events that satisfy one or more of the jet patch triggers.  If so, the embedded events are then passed through the full reconstruction and analysis routines that are used for the data.  The intermediate parton and particle records from \textsc{Pythia} are saved for all generated events, including those that fail the trigger simulation, to facilitate the study of potential bias effects.

\subsection{\textsc{PYTHIA} tune}

QCD events were generated using \textsc{Pythia} version 6.4.28 \cite{pythia2006} and the Perugia 2012 tune \cite{perugia2010,PerugiaUpdate}.  This combination overestimates the inclusive $\pi^{\pm}$ yields by up to 30\% for $p_T<3$ GeV/$c$, when compared to the previously published STAR measurements at $\sqrt{s} = 200$ GeV \cite{starpipm2006,starpipm2012}.  To compensate, the PARP(90) parameter in \textsc{Pythia} was reduced from 0.24 to 0.213.  PARP(90) controls the energy dependence of the low-$p_T$ cut-off for the UE generation process.  After this change, the simulated inclusive $\pi^{\pm}$ yields at $p_T<3$ GeV/$c$ match the experimentally measured cross sections within 10\%. The full $pp$ embedding sample, consisting of tuned \textsc{Pythia} + \textsc{Geant} simulated events embedded into zero-bias data, was then generated.  Jets were then reconstructed from the simulated detector responses using the same anti-$k_T$ algorithm with $R=0.5$ as was used to reconstruct jets in the data.  The simulation provides an excellent description of many jet-related quantities, as shown in the next subsection.  However, it slightly overestimates the rate of UE production seen in the data.  Systematic uncertainties to account for this mismatch are discussed in Sect.~\@ \ref{sect:Asymmetries}.

\subsection{Comparisons between data and simulation}

\begin{figure}[tb]
\includegraphics[width=\columnwidth]{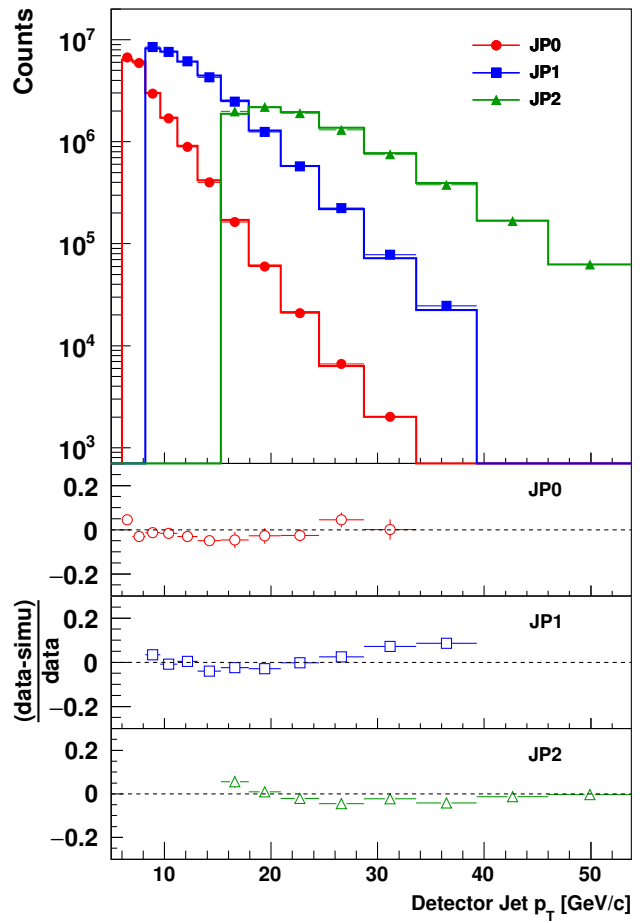}
\caption{\label{fig:Counts} The upper panel shows the jet yields \textit{vs}.\@ detector jet $p_T$ in data and simulation for each of the three trigger categories.  The points show the data, and the histograms show the simulation.  The lower three panels show the relative differences between data and simulation.
}
\end{figure}

\begin{figure}[tb]
\includegraphics[width=\columnwidth]{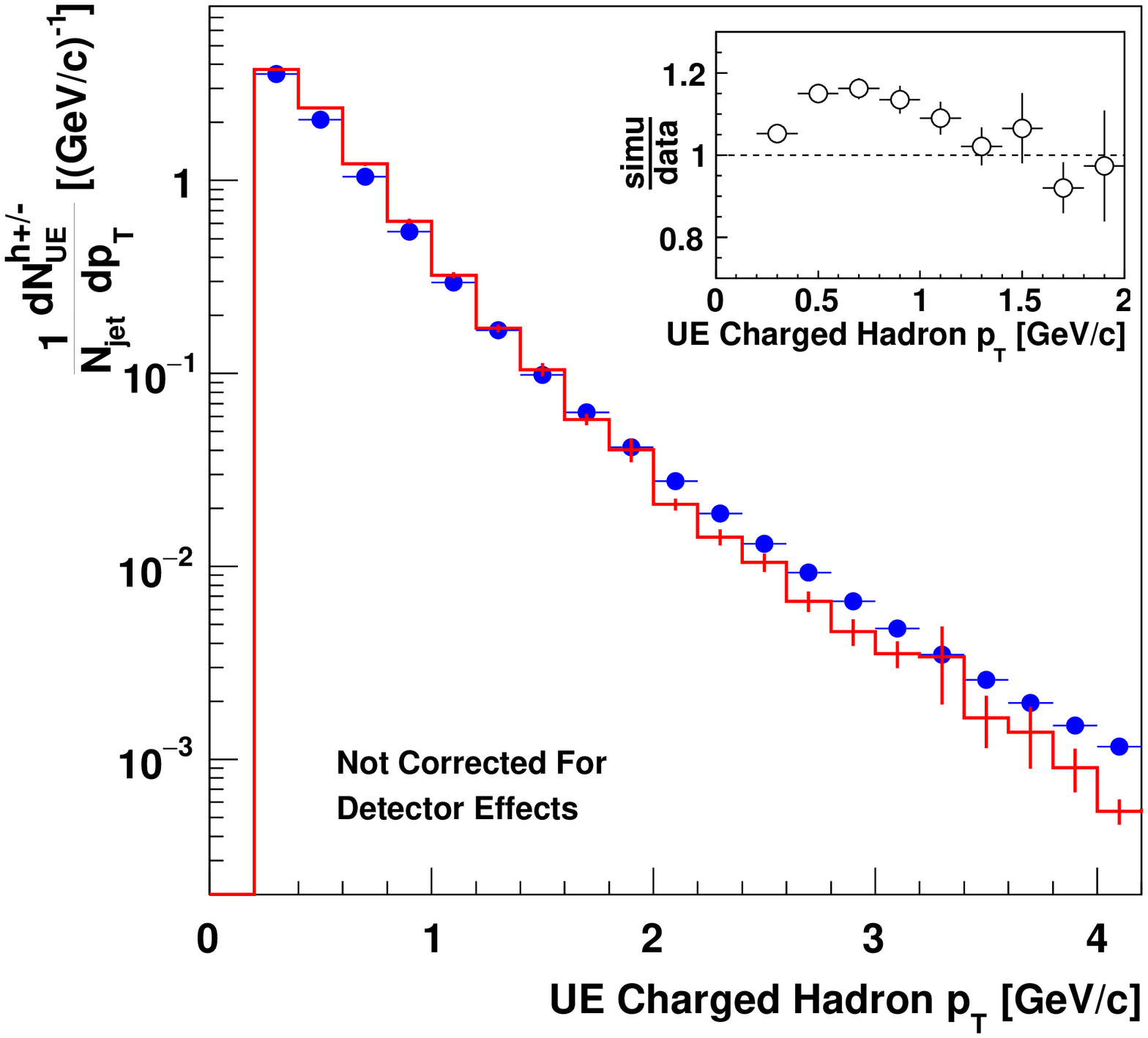}
\caption{\label{fig:UE_hadrons} Distributions of the charged hadrons within the off-axis cones as a function of hadron $p_T$.  The blue points show the data, and the red histogram shows the simulation.  The inset shows the ratio of simulation to data for $p_T < 2$ GeV/$c$.}
\end{figure}

Extensive comparisons of dijet and inclusive jet observables in the data and embedded simulation samples have been performed to ensure the simulation successfully reproduces the data. For the subset of these comparisons shown in Figs.\@ \ref{fig:Counts}--\ref{fig:UE_integrated}, the UE subtraction has been applied in the same way to the detector jets in data and simulation.  However, only raw, detector-level quantities are plotted, uncorrected for acceptance, efficiency, or resolution effects.  When the statistical uncertainties are not visible, they are smaller than the data points.

\subsubsection{Comparison of jet and dijet observables}

Figure \ref{fig:Counts} shows the jet yields \textit{vs}.\@ $p_T$ in the data and embedding for the three trigger categories. There is an excellent match between data and embedding for all three categories. 

\begin{figure}[tb]
\includegraphics[width=\columnwidth]{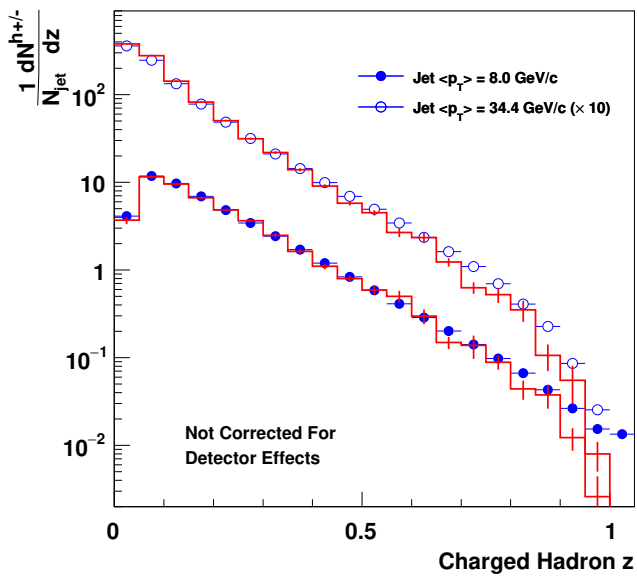}
\caption{\label{fig:z_dist} Distributions of the charged hadrons within the jets as a function of the hadron longitudinal momentum fraction, $z$, for two typical detector jet $p_T$ bins.  The blue points show the data, and the red histograms show the simulation.
}
\end{figure}

\begin{figure}[tb]
\includegraphics[width=\columnwidth]{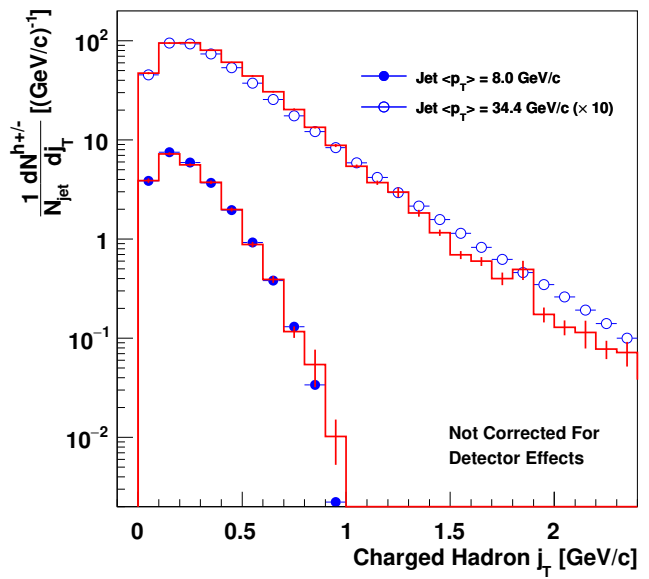}
\caption{\label{fig:jT_dist} Distributions of the charged hadrons within the jets as a function of the hadron momentum transverse to the thrust axis, $j_T$, for two typical detector jet $p_T$ bins.  The blue points show the data, and the red histograms show the simulation.
}
\end{figure}

Distributions of the charged hadrons within the jets are shown for data and simulation as functions of the hadron longitudinal momentum fraction, $z = p_{hadron}/p_{jet}$, and momentum transverse to the jet thrust axis, $j_T$, in Figs.\@ \ref{fig:z_dist} and \ref{fig:jT_dist}.  The distributions are shown for two representative detector jet $p_T$ bins, $7.0-8.2$~GeV/$c$ and $28.7-33.6$~GeV/$c$, which correspond to corrected mean jet $p_T$ values of 8.0 and 34.4~GeV/$c$ (see Sect.\@ \ref{subsect:scale_corrections}).

\begin{figure}[tb]
\includegraphics[width=\columnwidth]{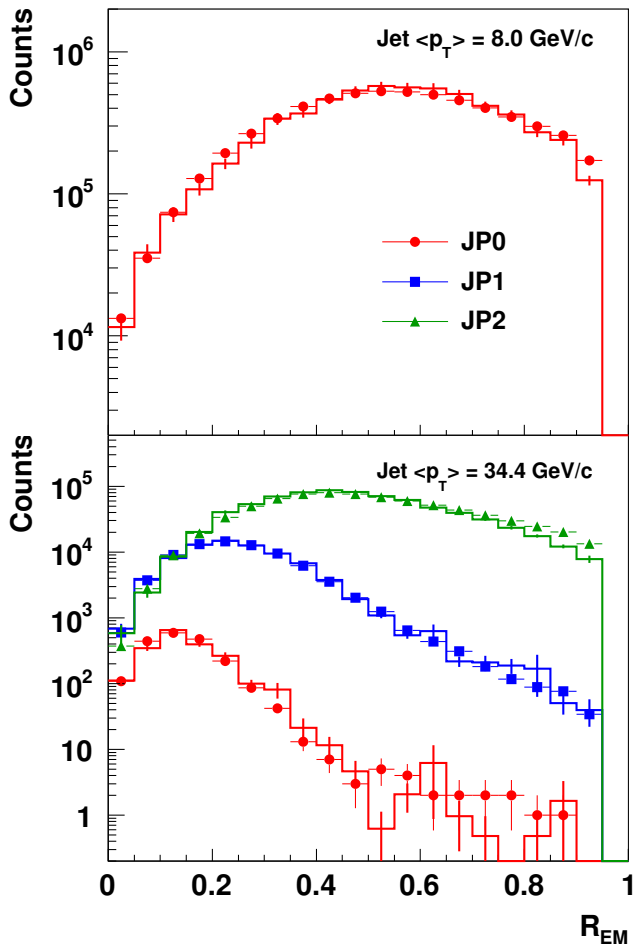}
\caption{\label{fig:R_EM} The upper panel shows the electromagnetic fraction distribution, $R_{EM}$, for jets in a low-$p_T$ bin.  The bias in favor of large $R_{EM}$, driven by the proximity to the JP0 threshold, is clear.  The lower panel shows the $R_{EM}$ distributions in a higher $p_T$ bin for each of the three jet categories.  For this case, JP0 and JP1 jets have a bias in favor of small $R_{EM}$, as discussed in the text.  The points show the data, and the histograms show the simulation.}
\end{figure}

An alternative view of the fragmentation and hadronization process is shown in Fig.\@ \ref{fig:R_EM}, which illustrates the $R_{EM}$ distributions for the same two jet $p_T$ bins.  The upper panel shows detector jets with $\langle p_T \rangle = 8.0$~GeV/$c$.  Only JP0-triggered jets contribute to this low-$p_T$ bin and the reconstructed jet $p_T$ is relatively close to the JP0 threshold, $E_T = 5.4$~GeV.  This favors jets with a large electromagnetic fraction, since only the energy deposited in the EMCs is considered by the trigger.  The jets with small $R_{EM}$ values that nonetheless satisfied the trigger contain charged hadrons that deposited an unusually large fraction of their energy in the EMCs.  This picture is reversed in the lower panel, which shows detector jets with $\langle p_T \rangle = 34.4$~GeV/$c$.  Typically, the jets in this momentum region require a large fraction of their energy to be carried by charged hadrons to be categorized as JP0 or JP1, instead of JP2.  The exceptions that have large $R_{EM}$ fractions typically were near the $\phi$-boundary between non-overlapping jet patches and shared their electromagnetic energy between them.  The data and embedding distributions in Figs.\@ \ref{fig:z_dist}, \ref{fig:jT_dist}, and \ref{fig:R_EM} match quite well, indicating that the simulations provide a very good description of the jet substructure.

%\subsubsection{Dijet comparisons}

\begin{figure}[b]
\includegraphics[width=\columnwidth]
{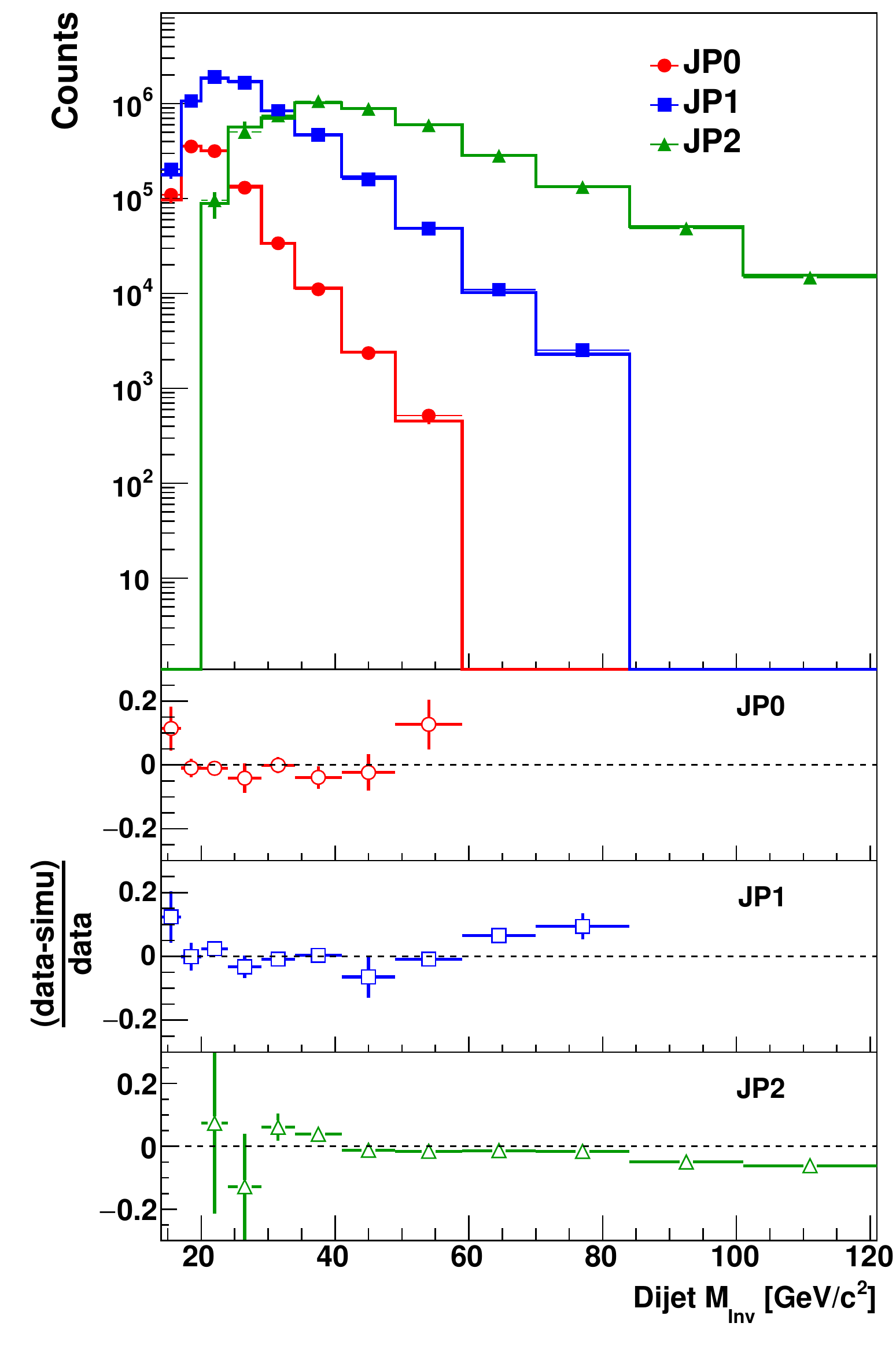}
\caption{\label{fig:DijetMass} Dijet yield \textit{vs}. invariant mass as measured at the detector level. In the top panel the data and simulations for the JP0, JP1 and JP2 trigger samples are represented by the points and histograms, respectively. The differences between data and simulation, normalized to the data yields, are shown for each of the trigger samples in the three bottom panels.}
\end{figure}

\begin{figure}[tb]
\includegraphics[width=\columnwidth]{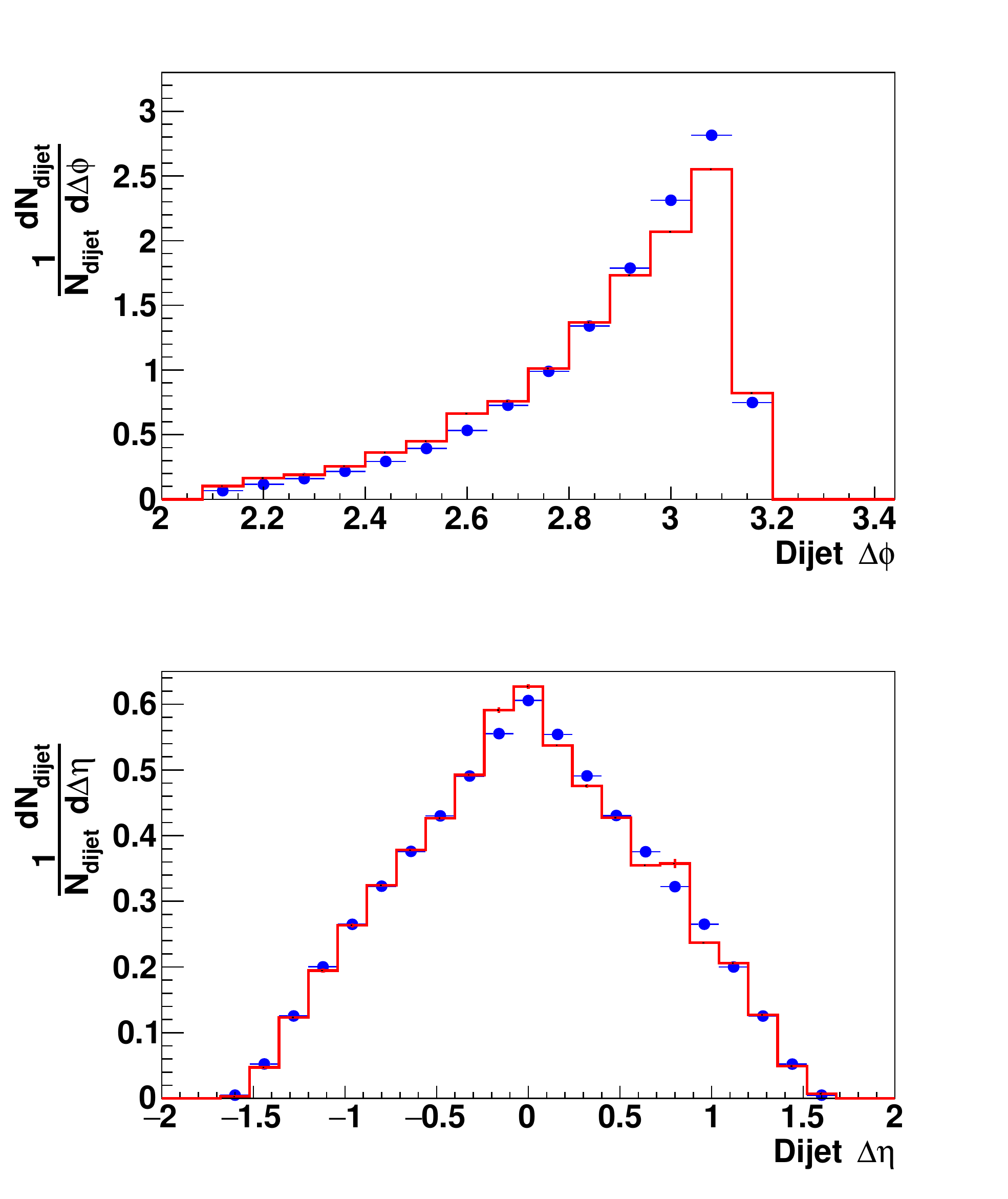}
\caption{\label{fig:DijetEtaPhi} Dijet opening angle $\Delta\phi$ (top) and $\Delta\eta$ (bottom) distributions. The blue points represent the data, and the red histograms the simulation.}
\end{figure}

Figure\@ \ref{fig:DijetMass} shows the comparison of data and simulation for the dijet yield as a function of invariant mass, $M_{inv}$.  The top panel in Fig.\@ \ref{fig:DijetEtaPhi} shows the opening angle $\Delta\phi$ and the bottom the $\Delta\eta$ of the dijet pair. No significant differences between the trigger samples were observed, therefore the independent trigger samples are combined, correctly accounting for run-time prescale in the simulation, for the $\Delta\phi$ and $\Delta\eta$ distributions. As with the inclusive yields, the agreement is excellent for the dijet observables. 

\subsubsection{Underlying event comparisons}

Figure \ref{fig:UE_hadrons} shows the distributions of charged hadrons within the off-axis cones in data and simulation as a function of hadron $p_T$.  The simulation provides a qualitative description of the observed UE hadrons.  But quantitatively it overestimates the UE production in the region $p_T \lesssim 1.3$~GeV/$c$, as illustrated in the inset.  The distributions of EMC tower $E_T$ values in the off-axis cones (not shown) also reveal an excess yield in the simulation at low $E_T$.

\begin{figure}[tbh]
\includegraphics[width=\columnwidth]
{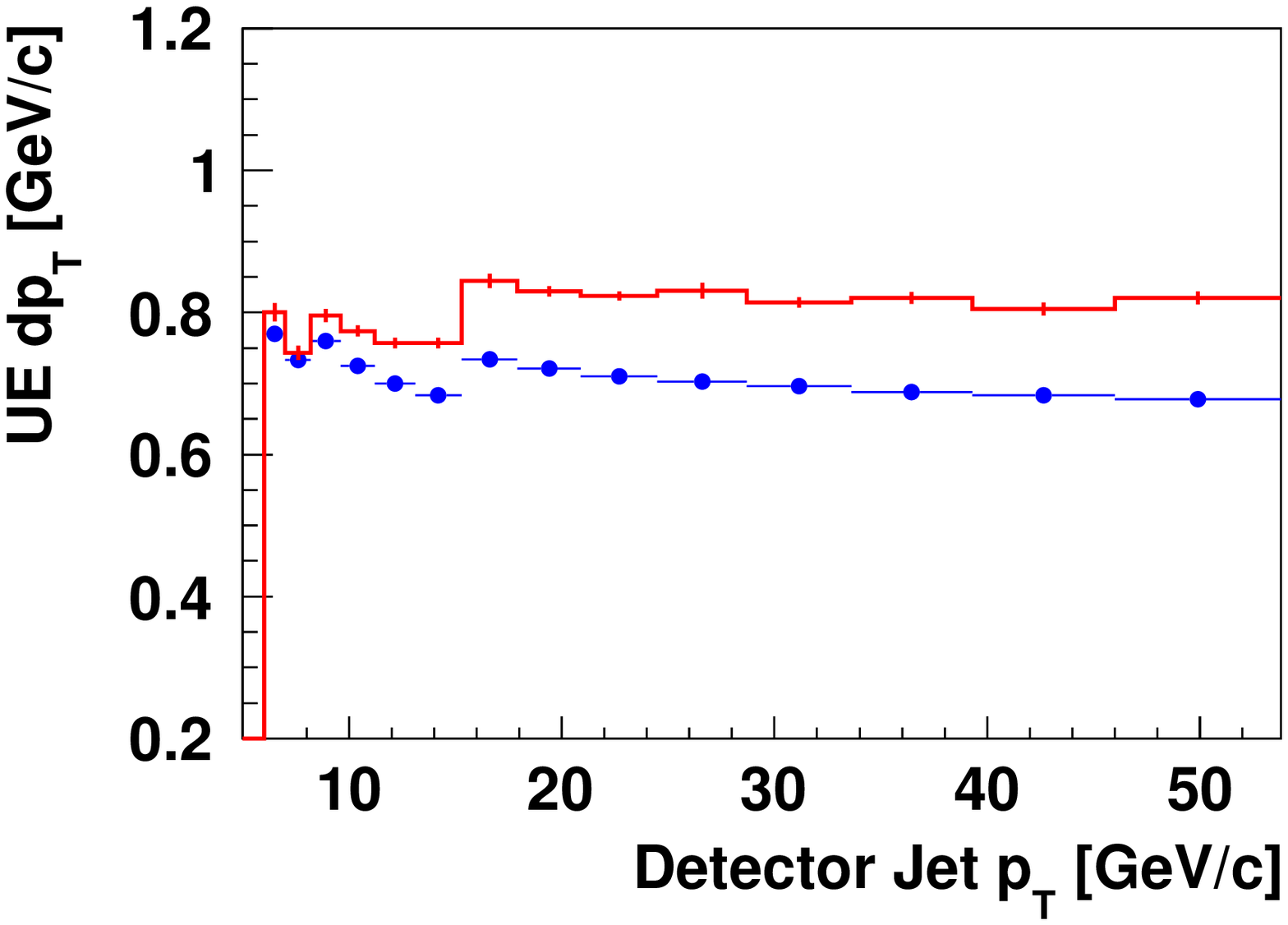}
\includegraphics[width=\columnwidth]
{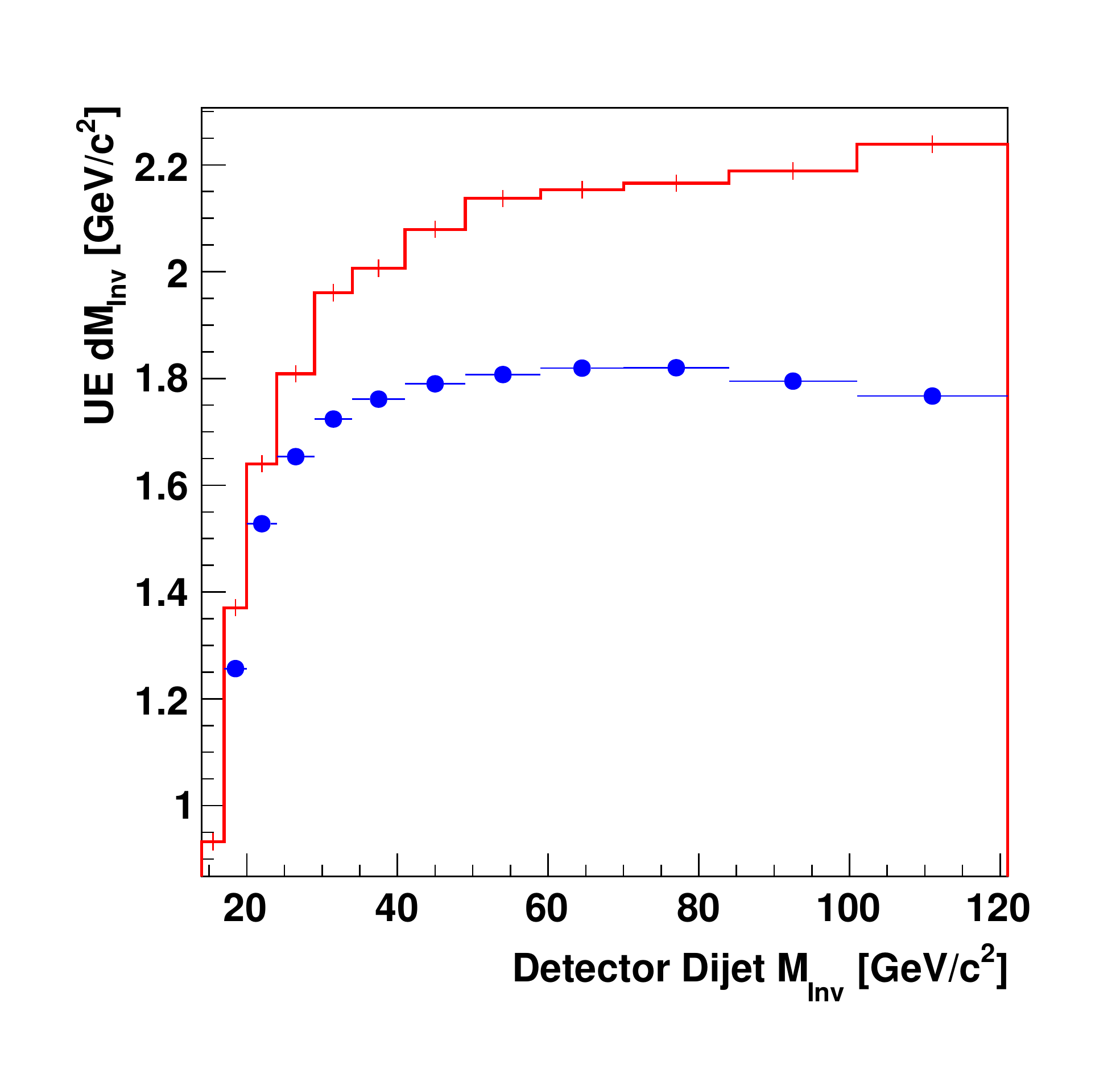}
\caption{\label{fig:UE_integrated}  The upper panel shows the mean underlying event correction to the jet transverse momentum, $dp_T$, as a function of detector jet $p_T$.  The lower panel shows the mean underlying event correction to the dijet mass, $dM_{inv}$, as a function of $M_{inv}$.  In both cases, the blue points show the data, and the red histogram shows the simulation.}
\end{figure}

The mean underlying event correction to the jet transverse momentum, $dp_T$, is shown in the upper panel of Fig.\@ \ref{fig:UE_integrated} as a function of detector jet $p_T$.  The discontinuities in the UE $dp_T$ distribution at 8.2 and 15.3~GeV/$c$, where the JP1 and JP2 event categories first contribute, originate from a trigger bias effect. As discussed in Sect.~\@\ref{sect:Experiment}, the STAR jet trigger is based on the energy observed in the EMCs. The UE present in an event serves to lower the effective trigger thresholds for the jets of interest, and hence increase the trigger efficiency at a given jet $p_T$.  This effect is maximal near the trigger turn-on points.
%As a result, the probability of a jet to register a JP1 or JP2 trigger decreases rapidly as the transverse energy ($E_T$) of the jet approaches the trigger $E_T$ threshold. In these regions of very low trigger efficiency the diffuse UE energy in the event can provide a substantial relative increase in trigger efficiency, leading to the enhancements at 8.2 and 15.3 GeV/$c$ in Figure \ref{fig:UE_integrated}. Naturally this effect is largest for those events with high UE event fluctuations. 
The lower panel of Fig.\@ \ref{fig:UE_integrated} shows a similar comparison of the UE corrections to the dijet mass, $dM_{inv}$, in data and simulation. In this case the enhancements at the trigger thresholds are less pronounced because only one of the two jets in the dijet pair is required to fulfill the trigger threshold requirement.%The diffuse UE energy present in an event lowers the effective hardware JP thresholds. Normally, this is a small effect. But close to the hardware threshold, where the absolute trigger efficiency is small, UE can provide a substantial relative increase in the trigger efficiency. This effect is largest for those events with the largest UE.

Overall, the UE yield discrepancy between data and simulation seen in Figs.\@ \ref{fig:UE_hadrons} and \ref{fig:UE_integrated} causes the simulation to predict 10\% to 20\% larger $dp_T$ and $dM_{inv}$ than seen in the data.  The implications of these differences are discussed in Sect.\@ \ref{sect:Asymmetries}.

\subsection{Parton and particle jets}\label{subsection:PartonParticleJet}
%Up to this point, the discussion has focused on detector jets.  Detector jets, formed from the charged tracks and EMC towers observed in the experiment, have the advantage that they are reconstructed the exact same way in data and simulation.  Furthermore, their properties are very similar, as illustrated in the previous subsection.However, detector jet properties are influenced by finite detector acceptance, efficiency, and resolution effects, so corrections are needed to compensate.%
Up to this point, the discussion has focused on jet properties as they are reconstructed at the detector level, both in data and simulation. These properties are influenced by finite detector acceptance, efficiency and resolution effects. The \textsc{Pythia} record for a simulated event affords the opportunity to relate these ``detector jets'' to more idealized jet objects. We do this in two ways. We reconstruct ``particle jets'' by running the anti-$k_T$ algorithm over the complete set of stable particles produced in the event. The same off-axis cone UE correction procedure used for detector jets is applied to particle jets. We also reconstruct ``parton jets'' by running the anti-$k_T$ algorithm on the hard-scattered partons from a given simulated \textsc{Pythia} event, including the initial-state and final-state radiation associated with the process, but excluding those partons from beam remnants and multiple parton interactions. In both cases, the jet finding algorithm and the input parameters to the algorithm are the same as used when reconstructing detector jets. 

Detector jets reconstructed from the embedded simulation detector responses can be matched to companion parton or particle jets. In this analysis, a parton or particle jet is considered to match a detector jet if the  distance between the jet thrust axes in $\eta-\phi$ space is $\sqrt{\Delta \eta^2 + \Delta \phi^2} < 0.5$.  If more than one parton or particle jet matches a given detector jet, we choose the closest one in $\eta-\phi$ space.  We then use the properties of the matched parton or particle jet when estimating corrections for the detector jets. The probability that a detector jet with $6.0 < p_T < 8.2$ GeV/$c$ has a matching particle jet is 98\%.  For $p_T > 8.2$ GeV/$c$, over 99\% of detector jets have a matching particle jet. The probability that a detector jet with $6.0 < p_T < 7.0$ GeV/$c$ has a matching parton jet is 85\%.  The probability increases rapidly with increasing detector jet $p_T$, reaching $>99\%$ for detector jet $p_T > 9.6$ GeV/$c$. Dijets are considered matched if each jet in the pair satisfies the inclusive jet matching criterion. The dijet detector-parton matching fractions range from 95\%--97$\%$ for $M_{inv}=14-17$ GeV$/c^2$ and quickly reach unity by $M_{inv}=24$ GeV$/c^2$. The improved matching fractions for the dijets compared to the inclusive jets is predominately due to the requirement to reconstruct two nearly back-to-back jets. This significantly cuts down on the reconstruction of fake jets and jets whose axis is badly reconstructed due to underlying event or background contributions.

As a further test of the UE subtraction procedure, we examine the difference in \textsc{Pythia} between the transverse momenta of UE-corrected particle jets and matched parton jets, $\delta p_T = p_{T,parton}-p_{T,particle}$.  At low $p_T$, where relatively little initial- and final-state radiation accompanies the hard scattering, we find $\langle \delta p_T \rangle \approx 0.1$ GeV/$c$.  This is a much smaller difference than was seen for low-$p_T$ jets in our previous $\sqrt{s} = 200$ GeV analysis, where no UE correction was implemented \cite{run9aLL2015}. At high $p_T$, where substantial gluon radiation often accompanies a hard scattering, the off-axis cones capture a small amount of initial- and final-state radiation, in addition to the UE, so the UE-corrected particle jet transverse momenta underestimate their matched parton jets by $2-3$\%.

\section{Longitudinal double-spin asymmetries}\label{sect:Asymmetries}

The longitudinal double-spin asymmetry $A_{LL}$ is defined as the difference of cross sections when the two beams have the same and opposite helicities divided by their sum:
\begin{equation}
A_{LL} = \frac{\sigma^{++}-\sigma^{+-}}{\sigma^{++}+\sigma^{+-}} .
\end{equation}
In this analysis, $A_{LL}$ is calculated as:
\begin{equation}
A_{LL} = \frac{\sum_{runs} P_YP_B(N^{++} - r N^{+-})}{\sum_{runs} P_Y^2P_B^2(N^{++}+ r N^{+-})},
\end{equation}
where $N^{++}$($N^{+-}$) is the number of jets or dijets observed in a given run with the same (opposite) helicity beams, $P_Y$ and $P_B$ are the beam polarizations for the run, and the relative luminosity, $r=\mathcal{L}^{++}/\mathcal{L}^{+-}$, is the ratio of the luminosities for same and opposite helicity beams during the run. The beam polarizations and relative luminosities were treated as constant over the duration of a run. This was motivated by the fact that runs typically lasted $10-40$ minutes, a short time period compared to changes in these beam properties. 

\subsection{Beam polarization}
The CNI polarization measurements were performed at the beginning of each fill, at several hour increments during the fill, and at the end of each fill.  The RHIC polarimetry group uses the results to determine the initial beam polarization at the beginning of the fill, $P_0$, and the polarization decay rate, $dP/dt$ \cite{run12polresult}. For each run, the polarization is taken to be the interpolated value at the midpoint of the run.
%The interpolated polarization value is a good approximation for both the JP0 and JP1 samples, despite for the time dependent JP0 and JP1 trigger rates. 
Following the guidance from the RHIC polarimetry group \cite{run12polresult}, the systematic uncertainty on the product of the two beam polarizations $P_YP_B$ is 6.6\% for the data used here.  This is a common overall scale uncertainty for the final inclusive jet and dijet $A_{LL}$ results.

\subsection{Relative luminosity}\label{subsect:RelLumi}

The relative luminosity for each run is calculated using scalers that counted the number of VPD coincidences and VPD east and west singles bunch-by-bunch.  The observed event counts for each bunch are corrected for accidental and multiple coincidences \cite{rellumcdf2000}.  The corrected VPD coincidence yields, summed over all bunches in the run with the same spin combination (++, +$-$, $-$+, $-$$-$), are then used to calculate the relative luminosity $r$ for that run. The values of the relative luminosity vary fill-by-fill from 0.9 to 1.1, depending on the sequence of beam helicities used.  Only very small variations are observed within fills.

To estimate the systematic uncertainty in the relative luminosity calculation, the ratios obtained from the corrected VPD coincidence yields are compared run-by-run to similar ratios calculated using the corrected VPD east singles or west singles, and to the ratios calculated using the corrected number of ZDC coincidences, east singles, or west singles.  A wide range of additional comparisons are made by considering alternative combinations of spin states, such as those appropriate to measure a parity-violating longitudinal single-spin asymmetry with the blue or yellow beam.  Following this study, a systematic uncertainty of $1.3 \times 10^{-4}$ is associated with $r$.

\subsection{Scale corrections and systematic errors}\label{subsect:scale_corrections}

$A_{LL}$ varies slowly and approximately linearly over the full kinematic range of the current measurements.  This makes it practical to implement a bin-by-bin unfolding technique to correct the inclusive jet $p_T$ and dijet $M_{inv}$ for detector resolution and efficiency effects. The matching conditions discussed in Sect.\@ \ref{subsection:PartonParticleJet} are implemented in the simulation and the average partonic level $p_T$ or $M_{inv}$ is determined for each detector bin. The calculated asymmetry for that bin is then plotted at the average partonic $p_T$ or $M_{inv}$ value.  This scale is chosen to facilitate a more direct comparison to the NLO pQCD theoretical predictions, which do not include effects from hadronization or underlying event contributions.
%Tables showing the corresponding scale values corrected to the particle level are available in Supplementary Material \cite{supplementary}.
The small, higher-order distortions from resolution and efficiency that remain are compensated as part of the trigger and reconstruction bias correction described below.

This type of correction requires an evaluation of the accuracy of the TPC track $p_T$ and EMC $E_T$ calibrations and efficiencies (labeled Hadron resp. and EM resp., respectively, in Tables~\ref{tab:jet_pT} and \ref{tab:dijet_M}). The effect of the systematic overestimate of the underlying event in the simulation on the jet $p_T$ and dijet $M_{inv}$, as well as the uncertainty in the \textsc{Pythia} tune, must also be quantified.  Tables \ref{tab:jet_pT} and \ref{tab:dijet_M} present the estimated corrections for the inclusive jet transverse momentum and dijet mass scales and their systematic uncertainties. The following sub-subsections discuss them in more detail.

\begin{table*}[tbh]
\caption{\label{tab:jet_pT} The corrections and systematic uncertainties assigning parton jet $p_T$ values to the detector-level inclusive jet $p_T$ bins.  The uncertainty quoted for $\delta p_T = \langle p_{T,parton} - p_{T,detector} \rangle$ is the contribution from the simulation statistics.  All values are in GeV/$c$.}
\begin{tabular}{c@{~~~}c@{~~~}c@{~~}|@{~~}c@{~~}cccc@{~~}|@{~~}c}
\hline\hline
   & Detector jet & & & & & & & Parton jet \\
Bin & $p_T$ range & $\langle p_T \rangle$ & $\delta p_T$ & ~Hadron resp.~ & ~EM resp.~ & ~UE syst.~ & ~Tune syst.~ & $p_T$ \\
\hline

%1 & $6.0-7.0$ & 6.48 & $0.54 \pm 0.06$ & 0.10 & 0.15 & 0.03 & 0.18 & $7.02 \pm 0.26$ \\
%2 & $7.0-8.2$ & 7.56 & $0.41 \pm 0.06$ & 0.16 & 0.16 & 0.01 & 0.18 & $7.97 \pm 0.30$ \\
%3 & $8.2-9.6$ & 8.86 & $1.04 \pm 0.06$ & 0.19 & 0.20 & 0.04 & 0.22 & $9.90 \pm 0.36$ \\
%4 & $9.6-11.2$ & 10.35 & $1.21 \pm 0.05$ & 0.21 & 0.22 & 0.05 & 0.25 & $11.56 \pm 0.40$ \\
%5 & $11.2-13.1$ & 12.07 & $1.30 \pm 0.05$ & 0.28 & 0.24 & 0.06 & 0.27 & $13.37 \pm 0.46$ \\
%6 & $13.1-15.3$ & 14.09 & $1.52 \pm 0.04$ & 0.34 & 0.26 & 0.07 & 0.24 & $15.61 \pm 0.50$ \\
%7 & $15.3-17.9$ & 16.52 & $2.47 \pm 0.06$ & 0.35 & 0.36 & 0.11 & 0.30 & $18.99 \pm 0.60$ \\
%8 & $17.9-20.9$ & 19.28 & $2.88 \pm 0.05$ & 0.39 & 0.42 & 0.11 & 0.23 & $22.17 \pm 0.63$ \\
%9 & $20.9-24.5$ & 22.52 & $3.14 \pm 0.05$ & 0.47 & 0.47 & 0.11 & 0.30 & $25.66 \pm 0.74$ \\
%10 & $24.5-28.7$ & 26.36 & $3.30 \pm 0.06$ & 0.60 & 0.52 & 0.13 & 0.21 & $29.65 \pm 0.83$ \\
%11 & $28.7-33.6$ & 30.81 & $3.56 \pm 0.07$ & 0.70 & 0.57 & 0.12 & 0.26 & $34.38 \pm 0.95$ \\
%12 & $33.6-39.3$ & 36.00 & $3.72 \pm 0.08$ & 0.82 & 0.64 & 0.13 & 0.22 & $39.7 \pm 1.1$ \\
%13 & $39.3-46.0$ & 42.06 & $4.26 \pm 0.09$ & 0.96 & 0.74 & 0.12 & 0.19 & $46.3 \pm 1.2$ \\
%14 & $46.0-53.8$ & 49.14 & $4.67 \pm 0.11$ & 1.11 & 0.85 & 0.14 & 0.49 & $53.8 \pm 1.5$ \\
~I1 & $6.0-7.0$ & 6.48 & $0.54 \pm 0.07$ & 0.10 & 0.15 & 0.03 & 0.18 & $7.02 \pm 0.26$ \\
~I2 & $7.0-8.2$ & 7.56 & $0.41 \pm 0.07$ & 0.16 & 0.16 & 0.01 & 0.18 & $7.97 \pm 0.30$ \\
~I3 & $8.2-9.6$ & 8.86 & $1.04 \pm 0.06$ & 0.19 & 0.20 & 0.04 & 0.22 & $9.90 \pm 0.36$ \\
~I4 & $9.6-11.2$ & 10.35 & $1.21 \pm 0.05$ & 0.21 & 0.22 & 0.05 & 0.25 & $11.56 \pm 0.40$ \\
~I5 & $11.2-13.1$ & 12.07 & $1.30 \pm 0.06$ & 0.28 & 0.24 & 0.06 & 0.27 & $13.37 \pm 0.47$ \\
~I6 & $13.1-15.3$ & 14.09 & $1.52 \pm 0.05$ & 0.34 & 0.26 & 0.07 & 0.24 & $15.61 \pm 0.50$ \\
~I7 & $15.3-17.9$ & 16.52 & $2.47 \pm 0.08$ & 0.35 & 0.36 & 0.11 & 0.30 & $18.99 \pm 0.60$ \\
~I8 & $17.9-20.9$ & 19.28 & $2.88 \pm 0.05$ & 0.39 & 0.42 & 0.11 & 0.23 & $22.17 \pm 0.63$ \\
~I9 & $20.9-24.5$ & 22.52 & $3.14 \pm 0.05$ & 0.47 & 0.47 & 0.11 & 0.30 & $25.66 \pm 0.74$ \\
I10 & $24.5-28.7$ & 26.36 & $3.30 \pm 0.06$ & 0.60 & 0.52 & 0.13 & 0.21 & $29.65 \pm 0.83$ \\
I11 & $28.7-33.6$ & 30.81 & $3.56 \pm 0.07$ & 0.70 & 0.57 & 0.12 & 0.26 & $34.38 \pm 0.95$ \\
I12 & $33.6-39.3$ & 36.00 & $3.72 \pm 0.08$ & 0.82 & 0.64 & 0.13 & 0.22 & $39.7 \pm 1.1$ \\
I13 & $39.3-46.0$ & 42.06 & $4.26 \pm 0.09$ & 0.96 & 0.74 & 0.12 & 0.19 & $46.3 \pm 1.2$ \\
I14 & $46.0-53.8$ & 49.14 & $4.67 \pm 0.11$ & 1.11 & 0.85 & 0.14 & 0.49 & $53.8 \pm 1.5$ \\
\hline\hline
\end{tabular}
\end{table*}

\begin{table*}[tbh]
\caption{\label{tab:dijet_M} The corrections and systematic uncertainties assigning parton dijet $M_{inv}$ to the detector-level dijet $M_{inv}$ bins.  The four topology groups are described in detail in Table \ref{table:topological}.  The uncertainty quoted for $\delta M_{inv} = \langle M_{inv,parton} - M_{inv,detector} \rangle$ is the contribution from the simulation statistics.  All values are in GeV/$c^2$.}
\begin{tabular}{c@{~}c@{~}c@{~}|@{~}c@{~~}c@{~~}c@{~~}c@{~~}c@{~}|@{~}c}
\hline\hline
   & Detector dijet & & & & & & & Parton dijet \\
Bin & $M_{inv}$ range & $\langle M_{inv} \rangle$ & $\delta M_{inv}$ & Hadron resp. & EM resp. & UE syst. & Tune syst. & $M_{inv}$ \\
\hline
\multicolumn{9}{c}{Topology A:~~~  Forward-Forward Dijets} \\
\hline
A1 & $14-17$ & 15.88 & $3.16 \pm 0.28$ & 0.41 & 0.30 & 0.16 & 0.58 & $19.04 \pm 0.82$ \\
A2 & $17-20$ & 18.48 & $3.14 \pm 0.22$ & 0.58 & 0.34 & 0.14 & 0.63 & $21.62 \pm 0.96$ \\
A3 & $20-24$ & 21.93 & $4.45 \pm 0.17$ & 0.59 & 0.39 & 0.12 & 0.78 & $26.38 \pm 1.08$ \\
A4 & $24-29$ & 26.34 & $5.96 \pm 0.23$ & 0.73 & 0.49 & 0.18 & 0.70 & $32.30 \pm 1.16$ \\
A5 & $29-34$ & 31.35 & $7.07 \pm 0.24$ & 0.85 & 0.57 & 0.24 & 0.52 & $38.42 \pm 1.20$ \\
A6 & $34-41$ & 37.19 & $7.99 \pm 0.23$ & 1.17 & 0.71 & 0.15 & 0.50 & $45.18 \pm 1.48$ \\
A7 & $41-49$ & 44.55 & $8.87 \pm 0.26$ & 1.17 & 0.82 & 0.27 & 0.48 & $53.42 \pm 1.55$ \\
A8 & $49-59$ & 53.23 & $10.29 \pm 0.27$ & 1.43 & 0.94 & 0.28 & 0.42 & $63.52 \pm 1.80$ \\
A9 & $59-70$ & 63.58 & $11.97 \pm 0.34$ & 1.81 & 1.09 & 0.39 & 0.40 & $75.55 \pm 2.21$ \\
A10 & $70-84$ & 75.49 & $13.63 \pm 0.42$ & 1.90 & 1.26 & 0.28 & 0.48 & $89.12 \pm 2.38$ \\
\hline
\multicolumn{9}{c}{Topology B:~~~  Forward-Central Dijets} \\
\hline
B1 & $14-17$ & 16.01 & $2.79 \pm 0.28$ & 0.44 & 0.31 & 0.09 & 0.53 & $18.80 \pm 0.81$ \\
B2 & $17-20$ & 18.52 & $3.28 \pm 0.16$ & 0.54 & 0.35 & 0.10 & 0.66 & $21.80 \pm 0.94$ \\
B3 & $20-24$ & 21.94 & $4.43 \pm 0.12$ & 0.73 & 0.40 & 0.11 & 0.69 & $26.37 \pm 1.09$ \\
B4 & $24-29$ & 26.34 & $5.90 \pm 0.14$ & 0.71 & 0.49 & 0.16 & 0.61 & $32.24 \pm 1.07$ \\
B5 & $29-34$ & 31.36 & $7.06 \pm 0.17$ & 0.97 & 0.60 & 0.26 & 0.55 & $38.42 \pm 1.31$ \\
B6 & $34-41$ & 37.22 & $8.61 \pm 0.16$ & 1.04 & 0.72 & 0.27 & 0.53 & $45.83 \pm 1.41$ \\
B7 & $41-49$ & 44.58 & $9.56 \pm 0.16$ & 1.30 & 0.85 & 0.27 & 0.57 & $54.14 \pm 1.69$ \\
B8 & $49-59$ & 53.30 & $10.87 \pm 0.18$ & 1.44 & 0.98 & 0.33 & 0.42 & $64.17 \pm 1.83$ \\
B9 & $59-70$ & 63.67 & $12.39 \pm 0.24$ & 1.72 & 1.12 & 0.30 & 0.39 & $76.06 \pm 2.13$ \\
B10 & $70-84$ & 75.67 & $14.14 \pm 0.27$ & 2.02 & 1.30 & 0.38 & 0.39 & $89.81 \pm 2.48$ \\
B11 & $84-101$ & 90.68 & $17.24 \pm 0.34$ & 2.36 & 1.53 & 0.46 & 0.33 & $107.92 \pm 2.89$ \\
\hline
\multicolumn{9}{c}{Topology C:~~~  Central-Central Dijets} \\
\hline
C1 & $14-17$ & 15.89 & $3.92 \pm 0.34$ & 0.33 & 0.32 & 0.16 & 0.56 & $19.81 \pm 0.81$ \\
C2 & $17-20$ & 18.49 & $3.52 \pm 0.24$ & 0.50 & 0.34 & 0.05 & 0.91 & $22.01 \pm 1.12$ \\
C3 & $20-24$ & 21.93 & $4.23 \pm 0.24$ & 0.73 & 0.41 & 0.08 & 0.63 & $26.16 \pm 1.08$ \\
C4 & $24-29$ & 26.34 & $6.36 \pm 0.26$ & 0.64 & 0.50 & 0.09 & 0.81 & $32.70 \pm 1.19$ \\
C5 & $29-34$ & 31.36 & $7.35 \pm 0.29$ & 1.05 & 0.61 & 0.21 & 0.43 & $38.71 \pm 1.34$ \\
C6 & $34-41$ & 37.22 & $8.79 \pm 0.28$ & 0.90 & 0.73 & 0.35 & 0.56 & $46.01 \pm 1.36$ \\
C7 & $41-49$ & 44.57 & $9.32 \pm 0.32$ & 1.35 & 0.86 & 0.40 & 0.57 & $53.89 \pm 1.78$ \\
C8 & $49-59$ & 53.31 & $11.44 \pm 0.35$ & 1.25 & 0.99 & 0.36 & 0.48 & $64.75 \pm 1.74$ \\
C9 & $59-70$ & 63.65 & $13.50 \pm 0.39$ & 1.84 & 1.14 & 0.37 & 0.45 & $77.15 \pm 2.27$ \\
C10 & $70-84$ & 75.70 & $15.44 \pm 0.49$ & 2.06 & 1.32 & 0.33 & 0.58 & $91.14 \pm 2.58$ \\
\hline
\multicolumn{9}{c}{Topology D:~~~  Forward-Backward Dijets} \\
\hline
D1 & $14-17$ & 16.34 & $4.20 \pm 0.60$ & 0.90 & 0.32 & -0.23 & 0.56 & $20.54 \pm 2.14$ \\
D2 & $17-20$ & 18.68 & $3.41 \pm 0.34$ & 0.57 & 0.35 & 0.01 & 0.91 & $22.09 \pm 0.95$ \\
D3 & $20-24$ & 21.97 & $4.34 \pm 0.23$ & 0.59 & 0.40 & 0.10 & 0.63 & $26.31 \pm 0.98$ \\
D4 & $24-29$ & 26.37 & $5.35 \pm 0.24$ & 0.72 & 0.47 & 0.12 & 0.81 & $31.72 \pm 1.22$ \\
D5 & $29-34$ & 31.36 & $7.12 \pm 0.32$ & 0.84 & 0.57 & 0.18 & 0.43 & $38.48 \pm 1.31$ \\
D6 & $34-41$ & 37.25 & $7.96 \pm 0.29$ & 1.17 & 0.70 & 0.22 & 0.56 & $45.21 \pm 1.55$ \\
D7 & $41-49$ & 44.65 & $9.50 \pm 0.33$ & 1.25 & 0.84 & 0.28 & 0.57 & $54.15 \pm 1.69$ \\
D8 & $49-59$ & 53.39 & $10.91 \pm 0.30$ & 1.49 & 0.97 & 0.35 & 0.48 & $64.30 \pm 1.92$ \\
D9 & $59-70$ & 63.72 & $12.94 \pm 0.36$ & 1.69 & 1.12 & 0.36 & 0.45 & $76.66 \pm 2.15$ \\
D10 & $70-84$ & 75.76 & $13.81 \pm 0.44$ & 2.14 & 1.29 & 0.34 & 0.58 & $89.57 \pm 2.62$ \\
D11 & $84-101$ & 90.82 & $16.04 \pm 0.55$ & 2.42 & 1.52 & 0.39 & 0.41 & $106.86 \pm 2.97$ \\ 
\hline\hline
\end{tabular}
\end{table*}

\subsubsection{Detector response uncertainties}

The shift in scale from the detector to the parton level depends on the accuracy of the TPC tracking efficiency implemented in the simulation. Studies of simulated $pp$ collisions at $\sqrt{s} = 200$ GeV suggested a 4$\%$ uncertainty on the accuracy of the tracking efficiency \cite{LiaoyuanThesis}. A more conservative estimate of the uncertainty, 5$\%$, is used here to reflect the reduction of tracking efficiency and increase in uncertainty that occurs at the higher luminosities in 510 GeV $pp$ collisions.

The effect of the tracking efficiency uncertainty on the scale correction is calculated by first randomly rejecting 5\% of all reconstructed TPC tracks in the embedded simulation sample, and then rerunning the jet finder. The differences between the jet $p_T$ and dijet $M_{inv}$ are taken as the systematic contributions due to tracking efficiency uncertainty.

There is an uncertainty associated with how well our \textsc{Geant} simulation models the energy deposited in the EMCs by hadrons \cite{MarciaET} that are either not detected by the TPC or are detected but deposit some of their energy outside of the tower pointed to by the track.  This contribution to the systematic error varies from 1.5\% of the jet $p_T$ at low $p_T$ to 2\% at high $p_T$.

The EMC gains are established using a combination of minimum-ionizing particles and identified electrons.  The calibration uncertainty for 2012 is estimated to be 3.8\%.  We apply this to the observed electromagnetic energy fractions $R_{EM}$, leading to scale uncertainties that range from 2.2\% at low $p_T$ to 1.7\% at high $p_T$.

In addition to the three effects discussed above, there are smaller ($<1$\%) contributions from the uncertainty in the EMC efficiency simulation and the TPC momentum calibration.

\subsubsection{Underlying event correction}
The full 10\% to 20\% difference between data and simulation for the underlying event $dp_T$ and $dM_{inv}$ corrections is taken as the systematic uncertainty. The uncertainty is calculated on a bin-by-bin basis for the jet and dijet distributions.

\subsubsection{\textsc{Pythia} tune variation}
A change in the parameters of the Perugia 2012 \textsc{Pythia} tune may, in turn, cause a shift in the average partonic jet $p_T$ and dijet $M_{inv}$ determined for the scale correction. The nature and size of the shift is studied by implementing several relevant variants \cite{perugia2010,PerugiaUpdate} of the Perugia 2012 tune in \textsc{Pythia} and recalculating the corrections.

The alternative tunes selected include the choice of $\alpha_s(\frac{1}{2}p_{\perp})$ and $\alpha_s(2p_{\perp})$ for higher (tune 371) and lower (tune 372) initial- and final-state radiation respectively, the modification to less color re-connections (tune 374), the increase in either longitudinal (376) or transverse (377) fragmentations, a switch to MSTW 2008 LO PDFs rather than CTEQ6L1 LO PDFs (378), and a set of Innsbruck hadronization parameters (383).

The corrections for alternative tune pairs (371,~372) and (376,~377), which relate to initial+final state radiation and fragmentation respectively, bracket those for the default tune. Therefore, half of the absolute difference of the pair is taken as its contribution to the tune systematic uncertainty. Together with the difference in scale shift from the remaining tunes, they are added in quadrature to construct the total \textsc{Pythia} tune systematic error.

\subsection{Asymmetry corrections and systematic errors}

\begin{figure}[bh]
\includegraphics[width=\columnwidth]{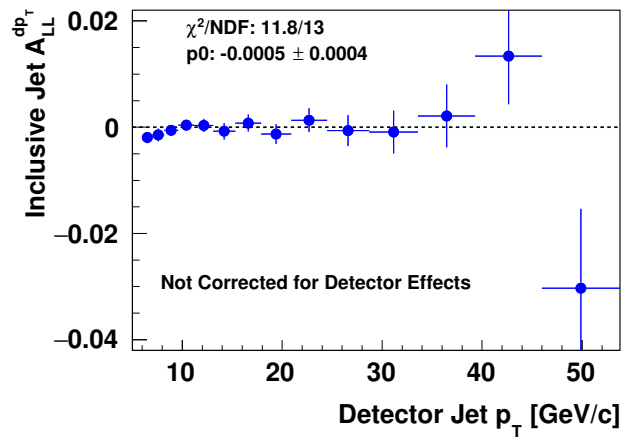}
\caption{\label{fig:UE_dpT_asym} Observed $A_{LL}^{dp_T}$ \textit{vs}. detector jet $p_T$.  Note that no corrections have been made for detector acceptance, efficiency, or resolution.}
\end{figure}

A broad range of systematic contributions to the measured $A_{LL}$ values are considered.  For low-$p_T$ jets the dominant contributions arise from UE and relative luminosity uncertainties, while trigger and reconstruction bias dominates at high $p_T$. The same trend is observed as function of $M_{inv}$ for the dijet sample. Several other effects are evaluated and found to be negligible compared to the statistical and leading systematic uncertainties.  Tables \ref{tab:jet_ALL_syst} and \ref{tab:dijet_ALL_syst} present the estimated corrections and systematic uncertainties for the inclusive jet and dijet $A_{LL}$ values, respectively.  The following sub-subsections describe these estimates.

\subsubsection{Underlying event contribution} \label{sect:UE_asym_syst}

Underlying event contributions both lower the effective JP trigger thresholds and increase the apparent energy of the reconstructed jets.  Thus, if the UE has a spin dependence, it can distort the measured dijet and inclusive jet $A_{LL}$ values.  To examine this possibility, we measured the longitudinal double-spin asymmetry of the underlying event contributions, $A_{LL}^{dp_T}$ and $A_{LL}^{dM_{inv}}$.  We define
\begin{equation}\label{Eq:dpT}
A_{LL}^{dp_T} = \frac{1}{P_YP_B} \frac{\langle dp_T \rangle^{++} - \langle dp_T \rangle^{+-}}{\langle dp_T \rangle^{++} + \langle dp_T \rangle^{+-}},
\end{equation}
where $\langle dp_T \rangle^{++}$ and $\langle dp_T \rangle^{+-}$ are the average underlying event corrections for same and opposite beam helicity combinations.  A similar definition is used for $A_{LL}^{dM_{inv}}$.  Figure \ref{fig:UE_dpT_asym} shows the observed $A_{LL}^{dp_T}$ as a function of detector jet $p_T$. The results in Fig.\@ \ref{fig:UE_dpT_asym} are not corrected for finite detector acceptance, efficiency, or resolution. However, these effects, which are independent of the beam spin combination, are expected to contribute similarly to the numerator and denominator in Eq.\@ (\ref{Eq:dpT}).  A constant fit finds $A_{LL}^{dp_T} = -0.0005 \pm 0.0004$, with $\chi^2 = 11.8$ for 13 degrees of freedom. A similar estimate was performed for each topology bin in the dijet analysis. The constant fit to the measured UE $A_{LL}^{dM_{inv}}$ for topology bin A finds $A_{LL}^{dM_{inv}}$ = $-0.0014 \pm 0.0017$, for bin B finds $A_{LL}^{dM_{inv}}$ = $0.0012 \pm 0.0011$, for bin C finds $A_{LL}^{dM_{inv}}$ = $-0.0035 \pm 0.0021$, and for bin D finds $A_{LL}^{dM_{inv}}$ = $0.0028 \pm 0.0015$, with associated $\chi^2$ per degrees of freedom values ranging from 0.5--1.2. Within the present statistics, $A_{LL}^{dp_T}$ and $A_{LL}^{dM_{inv}}$ are consistent with zero and independent of jet $p_T$ and dijet $M_{inv}$.

To estimate the possible systematic contribution that the UE can make to the measured inclusive jet (dijet) $A_{LL}$ values, we calculate the change in the cross section that would occur if the effective boundaries of our jet $p_T$ $(M_{inv})$ bins shift in a spin-dependent manner by an amount equal to the observed $dp_T$ $(dM_{inv})$ values multiplied by either the average UE asymmetry or the error, whichever is larger. For the inclusive jets the corresponding systematic uncertainties vary from $2.9\times10^{-4}$ for the lowest $p_T$ bin to $0.5\times10^{-4}$ for the highest $p_T$ bin. The dijet analysis follows a similar trend with errors of order $\approx1\times10^{-3}$ in the lower $M_{inv}$ bins that reduce to the level of $\approx2\times10^{-4}$ in the highest $M_{inv}$ bins. Since underlying event effects are expected to be independent of the hard scale and we use the overall average from the full data set to set the limit in the inclusive jet analysis, we treat these uncertainties as fully correlated. For the dijets, the errors are treated as fully correlated within a single topological bin.

\begin{table*}[tbh]
\caption{\label{tab:jet_ALL_syst} The corrections and systematic uncertainties in $A_{LL}$ for inclusive jet production.  In addition to the uncertainties enumerated here, there are two that are common to all the points, a shift uncertainty of $\pm 0.00022$ associated with the relative luminosity measurement and a scale uncertainty of $\pm 6.6\%$ associated with the beam polarization.}
\begin{tabular}{cc|c|cccc}
\hline\hline
   &  Jet $p_T$  &  &   \multicolumn{4}{c}{Trigger and Reconstruction Bias}  \\
~~Bin~~ & ~~(GeV/$c$)~~ & ~~UE syst.~~ & ~~Correction~~ & ~~PDF uncert.~~ & ~~Stat+vertex syst.~~ & ~~ Total syst.~~ \\
\hline

%1  &  ~7.02  &  0.00029  &  -0.00012  &  0.00013  &  0.00007  &  0.00015  \\
%2  &  ~7.97  &  0.00024  &  -0.00007  &  0.00037  &  0.00034  &  0.00050  \\
%3  &  ~9.90  &  0.00022  &  -0.00021  &  0.00007  &  0.00007  &  0.00010  \\
%4  &  11.56  &  0.00018  &  -0.00014  &  0.00007  &  0.00008  &  0.00011  \\
%5  &  13.37  &  0.00016  &  -0.00024  &  0.00008  &  0.00009  &  0.00012  \\
%6  &  15.61  &  0.00013  &  -0.00027  &  0.00009  &  0.00010  &  0.00013  \\
%7  &  18.99  &  0.00012  &  -0.00033  &  0.00011  &  0.00018  &  0.00021  \\
%8  &  22.17  &  0.00011  &  -0.00026  &  0.00019  &  0.00022  &  0.00029  \\
%9  &  25.66  &  0.00009  &  -0.00039  &  0.00013  &  0.00020  &  0.00024  \\
%10 &  29.65  &  0.00008  &  -0.00034  &  0.00020  &  0.00030  &  0.00036  \\
%11 &  34.38  &  0.00007  &  -0.00033  &  0.00025  &  0.00034  &  0.00042  \\
%12 &  39.7~  &  0.00006  &  -0.00004  &  0.00019  &  0.00045  &  0.00049  \\
%13 &  46.3~  &  0.00006  &   0.00042  &  0.00052  &  0.00066  &  0.00084  \\
%14 &  53.8~  &  0.00005  &   0.00011  &  0.00056  &  0.00089  &  0.00105  \\

~I1  &  ~7.02  &  0.00029  &  -0.00012  &  0.00013  &  0.00003  &  0.00013  \\
~I2  &  ~7.97  &  0.00024  &  -0.00007  &  0.00037  &  0.00032  &  0.00049  \\
~I3  &  ~9.90  &  0.00022  &  -0.00021  &  0.00007  &  0.00008  &  0.00011  \\
~I4  &  11.56  &  0.00018  &  -0.00014  &  0.00007  &  0.00004  &  0.00008  \\
~I5  &  13.37  &  0.00016  &  -0.00024  &  0.00008  &  0.00007  &  0.00011  \\
~I6  &  15.61  &  0.00013  &  -0.00027  &  0.00009  &  0.00013  &  0.00016  \\
~I7  &  18.99  &  0.00012  &  -0.00033  &  0.00011  &  0.00009  &  0.00014  \\
~I8  &  22.17  &  0.00011  &  -0.00026  &  0.00019  &  0.00013  &  0.00023  \\
~I9  &  25.66  &  0.00009  &  -0.00039  &  0.00013  &  0.00012  &  0.00018  \\
I10 &  29.65  &  0.00008  &  -0.00034  &  0.00020  &  0.00015  &  0.00025  \\
I11 &  34.38  &  0.00007  &  -0.00033  &  0.00025  &  0.00028  &  0.00038  \\
I12 &  39.7~  &  0.00006  &  -0.00004  &  0.00019  &  0.00028  &  0.00034  \\
I13 &  46.3~  &  0.00006  &   ~0.00042  &  0.00052  &  0.00045  &  0.00069  \\
I14 &  53.8~  &  0.00005  &   ~0.00011  &  0.00056  &  0.00059  &  0.00081  \\
\hline\hline
\end{tabular}
\end{table*}
%contents of this table have been updated (UE syst. column)
\begin{table*}[tbh]
\caption{\label{tab:dijet_ALL_syst} The corrections and systematic uncertainties in $A_{LL}$ for dijet production.  In addition to the uncertainties enumerated here, there are two that are common to all the points, a shift uncertainty of $\pm 0.00022$ associated with the relative luminosity measurement and a scale uncertainty of $\pm 6.6$\% associated with the beam polarization.  The four topology groups are described in detail in Table \ref{table:topological}.}
\begin{center}
\begin{tabular}{c@{~~~}c@{~~~}|@{~~~}c@{~~~}|@{~~~}c@{~~~}c@{~~~}c@{~~~}c}
\hline\hline
   &  Dijet $M_{inv}$  &  &   \multicolumn{4}{c}{Trigger and Reconstruction Bias}  \\
Bin & (GeV/$c^2$) & UE syst. & Model Correction & Model Error & Stat. Error & Total Error\\
\hline
\multicolumn{7}{c}{Topology A:~~~  Forward-Forward Dijets} \\
\hline
A1 & 19.04 & -0.00013 & 0.00057 & 0.00021 & 0.00002 & 0.00022\\
A2 & 21.62 & 0.00054 & 0.00071 & 0.00023 & 0.00004 & 0.00024\\
A3 & 26.38 & 0.00058 & 0.00119 & 0.00037 & 0.00003 & 0.00037\\
A4 & 32.30 & 0.00051 & 0.00109 & 0.00037 & 0.00001 & 0.00037\\
A5 & 38.42 & 0.00046 & 0.00184 & 0.00044 & 0.00001 & 0.00044\\
A6 & 45.18 & 0.00041 & 0.00143 & 0.00060 & 0.00001 & 0.00060\\
A7 & 53.42 & 0.00038 & 0.00246 & 0.00089 & 0.00001 & 0.00089\\
A8 & 63.52 & 0.00030 & 0.00077 & 0.00102 & 0.00004 & 0.00102\\
A9 & 75.55 & 0.00030 & 0.00256 & 0.00200 & 0.00005 & 0.00200\\
A10 & 89.12 & 0.00022 & 0.00658 & 0.00479 & 0.00002 & 0.00479\\
\hline
\multicolumn{7}{c}{Topology B:~~~  Forward-Central Dijets} \\
\hline
B1 & 18.80 & -0.00034 & 0.00058 & 0.00020 & 0.00002 & 0.00020\\
B2 & 21.80 & 0.00059 & 0.00085 & 0.00028 & 0.00001 & 0.00028\\
B3 & 26.37 & 0.00034 & 0.00113 & 0.00033 & 0.00002 & 0.00033\\
B4 & 32.24 & 0.00037 & 0.00101 & 0.00042 & 0.00001 & 0.00042\\
B5 & 38.42 & 0.00028 & 0.00172 & 0.00045 & 0.00003 & 0.00046\\
B6 & 45.83 & 0.00028 & 0.00115 & 0.00065 & 0.00001 & 0.00065\\
B7 & 54.14 & 0.00021 & 0.00112 & 0.00078 & 0.00001 & 0.00078\\
B8 & 64.17 & 0.00021 & 0.00245 & 0.00111 & 0.00003 & 0.00111\\
B9 & 76.06 & 0.00018 & 0.00278 & 0.00137 & 0.00001 & 0.00137\\
B10 & 89.81 & 0.00014 & 0.00519 & 0.00199 & 0.00001 & 0.00199\\
B11 & 107.92 & 0.00014 & 0.00732 & 0.00270 & 0.00003 & 0.00270\\
\hline
\multicolumn{7}{c}{Topology C:~~~  Central-Central Dijets} \\
\hline
C1 & 19.81 & -0.00015 & 0.00015 & 0.00019 & 0.00004 & 0.00019\\
C2 & 22.01 & 0.00059 & 0.00027 & 0.00063 & 0.00008 & 0.00064\\
C3 & 26.16 & 0.00070 & 0.00135 & 0.00041 & 0.00002 & 0.00041\\
C4 & 32.70 & 0.00075 & -0.00090 & 0.00057 & 0.00007 & 0.00057\\
C5 & 38.71 & 0.00078 & 0.00193 & 0.00069 & 0.00002 & 0.00069\\
C6 & 46.01 & 0.00048 & 0.00131 & 0.00058 & 0.00001 & 0.00058\\
C7 & 53.89 & 0.00048 & 0.00263 & 0.00117 & 0.00003 & 0.00117\\
C8 & 64.75 & 0.00035 & 0.00148 & 0.00143 & 0.00003 & 0.00143\\
C9 & 77.15 & 0.00035 & 0.00282 & 0.00185 & 0.00004 & 0.00185\\
C10 & 91.14 & 0.00035 & -0.00050 & 0.00732 & 0.00003 & 0.00732\\
\hline
\multicolumn{7}{c}{Topology D:~~~  Forward-Backward Dijets} \\
\hline
D1 & 20.54 & -0.00129 & 0.00067 & 0.00022 & 0.00003 & 0.00022\\
D2 & 22.09 & -0.00005 & 0.00065 & 0.00024 & 0.00003 & 0.00024\\
D3 & 26.31 & 0.00068 & 0.00086 & 0.00029 & 0.00004 & 0.00029\\
D4 & 31.72 & 0.00047 & 0.00132 & 0.00032 & 0.00002 & 0.00032\\
D5 & 38.48 & 0.00050 & 0.00113 & 0.00041 & 0.00001 & 0.00042\\
D6 & 45.21 & 0.00038 & 0.00151 & 0.00053 & 0.00007 & 0.00053\\
D7 & 54.15 & 0.00039 & 0.00171 & 0.00077 & 0.00002 & 0.00077\\
D8 & 64.30 & 0.00028 & 0.00296 & 0.00112 & 0.00001 & 0.00112\\
D9 & 76.66 & 0.00028 & 0.00482 & 0.00238 & 0.00002 & 0.00238\\
D10 & 89.57 & 0.00028 & 0.00273 & 0.00142 & 0.00001 & 0.00142\\
D11 & 106.86 & 0.00019 & 0.00178 & 0.00282 & 0.00003 & 0.00282\\
\hline\hline
\end{tabular}
\end{center}
\end{table*}

\subsubsection{Relative luminosity uncertainty}
The contribution to the total systematic uncertainty due to relative luminosity can be approximated as $\Delta A_{LL} = \frac{1}{2P_YP_B}\times \frac{\Delta r}{r}$. Taking $P_Y = 54 \%$, $P_B = 55\%$, and ${\Delta r}/{r}=1.3 \times 10^{-4}$, as calculated in Sect.\@ \ref{subsect:RelLumi}, this systematic uncertainty is estimated as $2.2 \times 10^{-4}$. It represents the possible offset of the $A_{LL}=0$ axis, and is common to all the measured inclusive and dijet asymmetries.

\subsubsection{Trigger bias and reconstruction uncertainty}\label{subsubsect:TrigBias}

\begin{figure}[tb]
\includegraphics*[bb=0 12 580 580,width=\columnwidth]{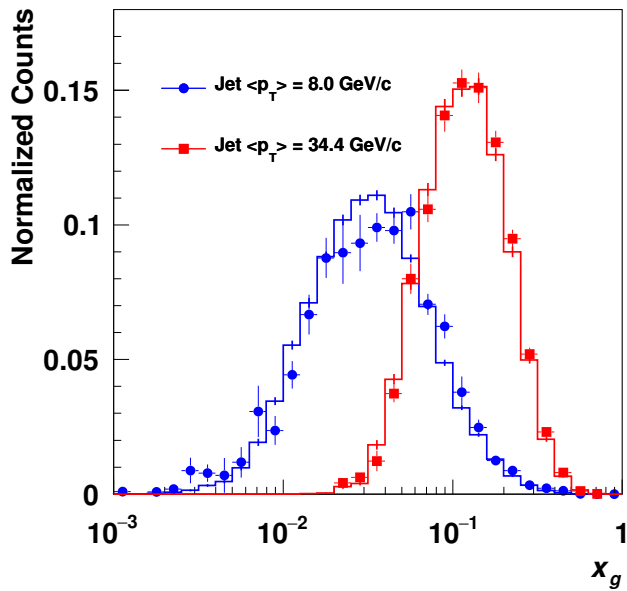}
\caption{\label{fig:inclusiveJetX} The $\hat{a}_{LL}$-weighted gluon $x$ distributions for two inclusive jet $p_T$ bins. The histograms represent all parton jets, independent of whether the jets satisfy the trigger and reconstruction requirements, while the points show the triggered detector jets.  Comparisons without the $\hat{a}_{LL}$ weight are qualitatively similar.}
\end{figure}

The jet matching calculations in Sect.\@ \ref{subsect:scale_corrections} closely align our jet measurements with those expected for unbiased parton jets, but the match is not perfect.  This is illustrated in Fig.\@ \ref{fig:inclusiveJetX}, which shows the sampled gluon $x$ distributions, weighted by the partonic asymmetry $\hat{a}_{LL}$ to indicate the region that is sensitive to $\Delta g(x)$.  At high jet $p_T$, the agreement is very good.  In contrast, at low jet $p_T$, there is a small but clear shift in $x$ between the unbiased distribution and the distribution that is sampled by the jets that are successfully triggered and reconstructed.  This difference arises from several trigger and reconstruction bias effects. For example, quark and gluon jets fragment differently, which can lead to different trigger efficiencies for gluon-gluon, quark-gluon, and quark-quark scattering processes. Detector and trigger resolutions might also distort the measured asymmetries. We utilize our embedding simulation to calculate a correction, and an associated uncertainty, to our measured $A_{LL}$ values to account for these trigger and reconstruction bias effects. At high collision rates, there is a small probability that low-$p_T$ jets will be assigned to the wrong vertex and, hence, mis-reconstructed.  We utilize the same embedding simulation to estimate the probability that the wrong vertex is found. 

The average luminosity during 2012 corresponds to $\approx$\,0.6 inelastic collisions per bunch crossing.  At this rate, there is a small probability that the highest quality vertex selected by the vertex finder was not the actual jet vertex.  We estimate this by comparing the reconstructed $z_\mathrm{vertex}$ in the simulation sample to the known position where the \textsc{Pythia} event was embedded.  For the two lowest-$p_T$ bins in the inclusive jet measurement, the wrong vertex is selected 15\% of the time.  This causes the jet kinematics, most especially the pseudorapidity, to be mis-reconstructed.  We assign a systematic uncertainty to $A_{LL}$ for these two bins, calculated by assuming events with the wrong vertex introduce a dilution of the true asymmetry. The probability to assign a jet to the wrong vertex is $\leq 4\%$ for the remaining inclusive jet $p_T$ bins, which makes this uncertainty negligible in all other cases. The probability to assign a dijet to the wrong vertex is less than $1\%$ for all invariant mass bins except for the forward-forward topology where the very lowest mass bins have a $5-7\%$ probability for a mismatched vertex. The quality of the vertex reconstruction is directly related to the number of tracks in the event and leads to the improved vertex matching found in the dijet sample. 

To estimate the remaining trigger and reconstruction biases, we compare the dijet and inclusive jet $A_{LL}$ values found by the simulation at the detector jet and parton jet levels.  To calculate $A_{LL}$ in the simulation, we weight each event by the product of the leading-order $2 \rightarrow 2$ partonic asymmetry $\hat{a}_{LL}$ and the ratio of polarized and unpolarized parton distribution functions of the two scattered partons,
\begin{equation}
w = \hat{a}_{LL}\frac{\Delta f_1(x_1,Q^2)\Delta f_2(x_2,Q^2)}{f_1(x_1,Q^2)f_2(x_2,Q^2)}.
\end{equation}
In principle, this requires knowledge of the polarized PDFs that we want to determine.  Lacking that, we calculate $A_{LL}$ using each of the 100 available equally probable replica sets that have been provided to span the range of polarized PDF uncertainties in NNPDFpol1.1 \cite{NNPDFpol}.  The average difference between the $A_{LL}$ found for all parton jets or dijets, including those from events that failed the trigger or detector jet reconstruction, and that for reconstructed detector jets or dijets is taken to be the correction for residual trigger and reconstuction bias effects.  The root-mean-square of the parton \textit{vs}.\@ detector $A_{LL}$ differences obtained with the 100 replica sets is assigned as a systematic uncertainty associated with our lack of knowledge of the true polarized PDFs.  This systematic is treated as fully correlated for all the $A_{LL}$ results.  We also assign a point-to-point systematic uncertainty to account for the finite statistics of the simulation.

Tables \ref{tab:jet_ALL_syst} and \ref{tab:dijet_ALL_syst} show that the trigger and reconstruction bias corrections are most significant at intermediate jet $p_T$ and dijet $M_{inv}$. For the lowest jet $p_T$ and dijet $M_{inv}$ values, which sample gluon $x$ values below the region that has been constrained by data in previous global analyses, the PDF uncertainties are larger than the calculated corrections.  At the highest jet $p_T$ and dijet $M_{inv}$ values, where the calculated bias is small, the statistical uncertainties dominate.  Nonetheless, the calculated corrections and their uncertainties are always small compared to the statistical uncertainties in the measured $A_{LL}$ results. 

\subsubsection{Other potential effects}

Residual transverse components of the beam polarization can distort the $A_{LL}$ measurement when coupled to the transverse double-spin asymmetry $A_{\Sigma}$~\cite{run6aLL2012}.  The residual transverse components of the beam polarization were monitored by the ZDC throughout the running period. Comparing the transverse asymmetries measured by the ZDC during transverse and longitudinal running periods, we find the transverse fractions of the total beam polarizations are approximately 5\% and 3\% for the blue and yellow beams, respectively. $A_{\Sigma}$ has not yet been measured in $\sqrt{s} = 510$ GeV collisions, but it has been measured to be less than 0.008 in the relevant $x$ range at $\sqrt{s} = 200$ GeV \cite{run6aLL2012}. If we take this as an upper limit, the contribution due to the residual transverse double-spin asymmetry is less than $3 \times 10^{-5}$, which is negligible compared to the other uncertainties.

Non-collision backgrounds can distort the $A_{LL}$ measurement if they satisfy our jet cuts. To estimate the non-collision background fraction, jets are reconstructed from the abort gaps in the same way as from the normal bunch crossings. Abort gaps are sequential bunch crossings where one of the beams has intentionally been left unfilled. Typically nine of the 120 bunch crossings in each beam were left unfilled during the 2012 running period. After cuts, the jet yield is reduced by four orders of magnitude relative to the yield from the normal bunch crossings, even though measures of the background rates in the zero-bias events are similar for both normal and abort gap crossings.  We conclude that non-collision backgrounds have a negligible impact on the $A_{LL}$ results.

At $\sqrt{s} = 510$ GeV, the parity-violating longitudinal single-spin asymmetry, $A_L$, is expected to be negligible compared to our current statistical precision.  Therefore, we examine the blue and yellow beam single-spin asymmetries $A_L^B$ and $A_L^Y$ as a consistency check of the relative luminosity calculations and an indicator of bunch-dependent collider instrumental effects.  As expected, the observed asymmetries are consistent with zero  and subsequently no corresponding systematic uncertainty is assigned.

\subsection{Correlations}\label{sect:correlations}

%The previous subsection indicates that a number of correlated systematic effects are present in the $A_{LL}$ measurements.%
Most of the dijet events contain at least one jet that satisfies the inclusive jet cuts.  This leads to significant statistical correlations between the dijet and inclusive jet results, as large as 0.21, when the dijet $M_{inv}$ is close to twice the inclusive jet $p_T$.  As noted in Sect.\@ \ref{sect:incl_jet_cuts}, a small fraction ($\simeq$\,5\%) of the events in the inclusive jet analysis contain two jets, both of which satisfy the inclusive jet cuts.  This produces statistical correlations between the inclusive jet asymmetry measurements that range from 0.01 for low-$p_T$ pairs of inclusive jet bins to 0.04 for high-$p_T$ pairs.  In contrast, there are no statistical correlations between the dijet asymmetry measurements.

There are also correlated point-to-point systematic effects, as discussed in the previous subsection, though they are typically smaller than the statistical correlations.  We treat the inclusive jet UE systematic uncertainty estimates in Table \ref{tab:jet_ALL_syst} as fully correlated because they are all derived from the same fit to the $A_{LL}^{dp_T}$ measurements shown in Fig.\@ \ref{fig:UE_dpT_asym}.  Similarly, we treat the dijet UE systematic uncertainty estimates in Table \ref{tab:dijet_ALL_syst} as fully correlated for each topology group because each dijet UE uncertainty is derived from the fit to the $A_{LL}^{dM_{inv}}$ measurements for that specific topology group.  The PDF systematic uncertainties in Tables \ref{tab:jet_ALL_syst} and \ref{tab:dijet_ALL_syst} are highly correlated for measurements that sample nearby $x$ values.  The correlation is weaker for measurements that sample more distant $x$ values.  To be conservative, we nonetheless treat all the PDF systematic uncertainties as fully correlated.  In all cases, the total point-to-point systematic correlations are estimated to be $<$\,0.06.  The full correlation matrix for the inclusive jet and dijet $A_{LL}$ measurements, including both the statistical and point-to-point systematic uncertainties, is given in the Appendix. 

There are two systematic uncertainties, relative luminosity and beam polarization, that are common to all of the measurements and are not included in the correlation matrix presented in the Appendix.  Note that the $\pm$\,6.6\% polarization scale uncertainty is common not just to the measurements here, but to all double-spin asymmetry measurements that are derived from 2012 RHIC data at $\sqrt{s}$ = 510 GeV, including those in \cite{PHENIXpp510,STARFMSpi0}.  Furthermore, a substantial fraction of the polarization scale uncertainty arises from uncertainty in the molecular hydrogen fraction in the hydrogen gas-jet target, and this uncertainty is correlated across several years of RHIC operation.  See \cite{run12polresult} for details.

\begin{table*}[tbh]
\caption{\label{tab:jet_ALL}$A_{LL}$ as a function of parton jet $p_T$ (in GeV/$c$) for inclusive jets with $|\eta| < 0.9$ in $\sqrt{s} = 510$ GeV $pp$ collisions.  There is an additional $\pm 6.6\%$ scale uncertainty from the beam polarization that is common to all the measurements.  The underlying event and relative luminosity systematics are fully correlated for all the points.}
\begin{tabular}{c@{~~~}c@{~~~}c@{~~~}c}
\hline\hline
Bin & Jet $p_T$ & $A_{LL}$ $\pm$ stat.\@ $\pm$ syst. & UE/RL syst. \\
\hline
%~1 & ~7.02\,$\pm$\,0.26  &  ~0.0002\,$\pm$\,0.0013\,$\pm$\,0.0001  &  0.00036  \\
%~2 & ~7.97\,$\pm$\,0.30  &  -0.0022\,$\pm$\,0.0014\,$\pm$\,0.0005  & 0.00033  \\
%~3 & ~9.90\,$\pm$\,0.36  &  ~0.0016\,$\pm$\,0.0010\,$\pm$\,0.0001  &  0.00031  \\
%~4 & 11.56\,$\pm$\,0.40  &  ~0.0005\,$\pm$\,0.0011\,$\pm$\,0.0001  &  0.00028  \\
%~5 & 13.37\,$\pm$\,0.46  &  ~0.0015\,$\pm$\,0.0013\,$\pm$\,0.0001  &  0.00027  \\
%~6 & 15.61\,$\pm$\,0.50  &  ~0.0029\,$\pm$\,0.0016\,$\pm$\,0.0001  &  0.00026  \\
%~7 & 18.99\,$\pm$\,0.60  &  ~0.0016\,$\pm$\,0.0016\,$\pm$\,0.0002  &  0.00025  \\
%~8 & 22.17\,$\pm$\,0.63  &  ~0.0044\,$\pm$\,0.0018\,$\pm$\,0.0003  &  0.00025  \\
%~9 & 25.66\,$\pm$\,0.74  &  ~0.0050\,$\pm$\,0.0021\,$\pm$\,0.0002  &  0.00024  \\
%10 & 29.65\,$\pm$\,0.83  &  ~0.0036\,$\pm$\,0.0027\,$\pm$\,0.0004  &  0.00023  \\
%11 & 34.38\,$\pm$\,0.95  &  ~0.0169\,$\pm$\,0.0037\,$\pm$\,0.0004  &  0.00023  \\
%12 & 39.7\,$\pm$\,1.1  &  -0.0049\,$\pm$\,0.0054\,$\pm$\,0.0005  &  0.00023  \\
%13 & 46.3\,$\pm$\,1.2  &  ~0.0122\,$\pm$\,0.0084\,$\pm$\,0.0008  &  0.00023  \\
%14 & 53.8\,$\pm$\,1.5  &  ~0.0018\,$\pm$\,0.0137\,$\pm$\,0.0011  &  0.00023  \\
~I1 & ~7.02\,$\pm$\,0.26  &  ~0.0002\,$\pm$\,0.0013\,$\pm$\,0.0001  &  0.00036  \\
~I2 & ~7.97\,$\pm$\,0.30  &  -0.0022\,$\pm$\,0.0014\,$\pm$\,0.0005  & 0.00033  \\
~I3 & ~9.90\,$\pm$\,0.36  &  ~0.0016\,$\pm$\,0.0010\,$\pm$\,0.0001  &  0.00031  \\
~I4 & 11.56\,$\pm$\,0.40  &  ~0.0005\,$\pm$\,0.0011\,$\pm$\,0.0001  &  0.00028  \\
~I5 & 13.37\,$\pm$\,0.47  &  ~0.0015\,$\pm$\,0.0013\,$\pm$\,0.0001  &  0.00027  \\
~I6 & 15.61\,$\pm$\,0.50  &  ~0.0029\,$\pm$\,0.0016\,$\pm$\,0.0002  &  0.00026  \\
~I7 & 18.99\,$\pm$\,0.60  &  ~0.0016\,$\pm$\,0.0016\,$\pm$\,0.0001  &  0.00025  \\
~I8 & 22.17\,$\pm$\,0.63  &  ~0.0044\,$\pm$\,0.0018\,$\pm$\,0.0002  &  0.00025  \\
~I9 & 25.66\,$\pm$\,0.74  &  ~0.0050\,$\pm$\,0.0021\,$\pm$\,0.0002  &  0.00024  \\
I10 & 29.65\,$\pm$\,0.83  &  ~0.0036\,$\pm$\,0.0027\,$\pm$\,0.0003  &  0.00023  \\
I11 & 34.38\,$\pm$\,0.95  &  ~0.0169\,$\pm$\,0.0037\,$\pm$\,0.0004  &  0.00023  \\
I12 & 39.7\,$\pm$\,1.1  &  -0.0049\,$\pm$\,0.0054\,$\pm$\,0.0003  &  0.00023  \\
I13 & 46.3\,$\pm$\,1.2  &  ~0.0122\,$\pm$\,0.0084\,$\pm$\,0.0007  &  0.00023  \\
I14 & 53.8\,$\pm$\,1.5  &  ~0.0018\,$\pm$\,0.0137\,$\pm$\,0.0008  &  0.00023  \\
\hline\hline
\end{tabular}
\end{table*}

\begin{table*}[tbh]
\caption{\label{tab:dijet_ALL}$A_{LL}$ as a function of parton dijet $M_{inv}$ (in GeV/$c^2$) in $\sqrt{s}= 510$ GeV $pp$ collisions.  There is an additional $\pm 6.6\%$ scale uncertainty from the beam polarization that is common to all the measurements.  The underlying event and relative luminosity systematics are fully correlated for each topology group.  The four topology groups are described in detail in Table \ref{table:topological}.}
\begin{tabular}{c@{~~~}c@{~~~}c@{~~~}c}
\hline\hline
Bin & Dijet $M_{inv}$ & $A_{LL}$ $\pm$ stat.\@ $\pm$ syst. & UE/RL syst. \\
\hline
\multicolumn{4}{c}{Topology A:~~~  Forward-Forward Dijets}\\
\hline
%A1 & ~18.65\,$\pm$\,1.037  &  ~9.85E-03\,$\pm$\,0.00501\,$\pm$\,0.00029  &  0.00022  \\
%A2 & ~21.24\,$\pm$\,1.196  &  ~9.96E-03\,$\pm$\,0.00471\,$\pm$\,0.00048  & 0.00026  \\
%A3 & ~25.77\,$\pm$\,1.362  &  -2.59E-04\,$\pm$\,0.00499\,$\pm$\,0.00029  &  0.00026  \\
%A4 & ~30.84\,$\pm$\,1.584  &  ~1.07E-02\,$\pm$\,0.00598\,$\pm$\,0.00032  &  0.00025  \\
%A5 & ~36.87\,$\pm$\,1.741  &  -8.64E-04\,$\pm$\,0.00625\,$\pm$\,0.00042  &  0.00025  \\
%A6 & ~43.61\,$\pm$\,2.098  &  ~5.64E-03\,$\pm$\,0.00796\,$\pm$\,0.00055  &  0.00024  \\
%A7 & ~52.12\,$\pm$\,2.239  &  ~3.38E-03\,$\pm$\,0.01060\,$\pm$\,0.00088  &  0.00024  \\
%A8 & ~62.44\,$\pm$\,2.701  &  ~3.71E-02\,$\pm$\,0.01622\,$\pm$\,0.00123  &  0.00023  \\
%A9 & ~74.76\,$\pm$\,3.259  &  -1.87E-02\,$\pm$\,0.02472\,$\pm$\,0.00189  &  0.00023  \\
%A10 & ~88.5\,$\pm$\,3.669  &  ~6.88E-02\,$\pm$\,0.04192\,$\pm$\,0.00343  &  0.00023  \\
A1 & ~19.04\,$\pm$\,0.82  &  -0.0128\,$\pm$\,0.0066\,$\pm$\,0.0002  &  0.00025  \\
A2 & ~21.62\,$\pm$\,0.96  &  0.0090\,$\pm$\,0.0052\,$\pm$\,0.0002  & 0.00058  \\
A3 & ~26.38\,$\pm$\,1.08  &  0.0079\,$\pm$\,0.0050\,$\pm$\,0.0004  &  0.00062  \\
A4 & ~32.30\,$\pm$\,1.16  &  -0.0012\,$\pm$\,0.0052\,$\pm$\,0.0004  &  0.00056  \\
A5 & ~38.42\,$\pm$\,1.20  &  0.0101\,$\pm$\,0.0061\,$\pm$\,0.0004  &  0.00051  \\
A6 & ~45.18\,$\pm$\,1.48  &  -0.0013\,$\pm$\,0.0064\,$\pm$\,0.0006  &  0.00046  \\
A7 & ~53.42\,$\pm$\,1.55  &  0.0048\,$\pm$\,0.0081\,$\pm$\,0.0009  &  0.00044  \\
A8 & ~63.52\,$\pm$\,1.80  &  0.0052\,$\pm$\,0.0108\,$\pm$\,0.0010  &  0.00037  \\
A9 & ~75.55\,$\pm$\,2.21  &  0.0363\,$\pm$\,0.0167\,$\pm$\,0.0020  &  0.00037  \\
A10 & ~89.12\,$\pm$\,2.38  &  -0.0218\,$\pm$\,0.0264\,$\pm$\,0.0048  &  0.00031  \\
\hline
\multicolumn{4}{c}{Topology B:~~~  Forward-Central Dijets}\\
\hline
%B1 & ~18.6\,$\pm$\,0.787  &  ~4.55E-03\,$\pm$\,0.00345\,$\pm$\,0.00019  &  0.00026  \\
%B2 & ~21.6\,$\pm$\,0.956  &  ~1.19E-03\,$\pm$\,0.00313\,$\pm$\,0.00020  & 0.00032  \\
%B3 & ~25.84\,$\pm$\,1.274  &  ~2.89E-03\,$\pm$\,0.00326\,$\pm$\,0.00030  &  0.00026  \\
%B4 & ~31.15\,$\pm$\,1.391  &  -5.65E-03\,$\pm$\,0.00390\,$\pm$\,0.00032  &  0.00026  \\
%B5 & ~36.95\,$\pm$\,1.657  &  ~4.15E-03\,$\pm$\,0.00401\,$\pm$\,0.00187  &  0.00025  \\
%B6 & ~43.82\,$\pm$\,1.974  &  ~1.12E-02\,$\pm$\,0.00493\,$\pm$\,0.00061  &  0.00025  \\
%B7 & ~52.48\,$\pm$\,2.399  &  ~3.27E-05\,$\pm$\,0.00636\,$\pm$\,0.00069  &  0.00023  \\
%B8 & ~62.63\,$\pm$\,2.754  &  ~1.41E-04\,$\pm$\,0.00943\,$\pm$\,0.00104  &  0.00023  \\
%B9 & ~74.75\,$\pm$\,3.229  &  -1.46E-02\,$\pm$\,0.01391\,$\pm$\,0.00134  &  0.00023  \\
%B10 & ~88.83\,$\pm$\,3.701  &  -1.14E-02\,$\pm$\,0.02283\,$\pm$\,0.00188  &  0.00023  \\
%B11 & 107.31\,$\pm$\,4.301  &  -6.46E-03\,$\pm$\,0.04057\,$\pm$\,0.00258  &  0.00023  \\
B1 & ~18.80\,$\pm$\,0.81  &  -0.0023\,$\pm$\,0.0053\,$\pm$\,0.0002  &  0.00040  \\
B2 & ~21.80\,$\pm$\,0.94  &  0.0041\,$\pm$\,0.0036\,$\pm$\,0.0003  & 0.00063  \\
B3 & ~26.37\,$\pm$\,1.09  &  0.0016\,$\pm$\,0.0033\,$\pm$\,0.0003  &  0.00041  \\
B4 & ~32.24\,$\pm$\,1.07  &  0.0029\,$\pm$\,0.0034\,$\pm$\,0.0004  &  0.00043  \\
B5 & ~38.42\,$\pm$\,1.31  &  -0.0063\,$\pm$\,0.0040\,$\pm$\,0.0005  &  0.00035  \\
B6 & ~45.83\,$\pm$\,1.41  &  0.0020\,$\pm$\,0.0041\,$\pm$\,0.0007  &  0.00036  \\
B7 & ~54.14\,$\pm$\,1.69  &  0.0128\,$\pm$\,0.0050\,$\pm$\,0.0008  &  0.00030  \\
B8 & ~64.17\,$\pm$\,1.83  &  -0.0022\,$\pm$\,0.0065\,$\pm$\,0.0011  &  0.00031  \\
B9 & ~76.06\,$\pm$\,2.13  &  -0.0010\,$\pm$\,0.0096\,$\pm$\,0.0014  &  0.00028  \\
B10 & ~89.81\,$\pm$\,2.48  &  -0.0160\,$\pm$\,0.0143\,$\pm$\,0.0020  &  0.00026  \\
B11 & 107.92\,$\pm$\,2.89  &  -0.0205\,$\pm$\,0.0242\,$\pm$\,0.0027  &  0.00026  \\
\hline
\multicolumn{4}{c}{Topology C:~~~  Central-Central Dijets}\\
\hline
%C1 & ~19.32\,$\pm$\,1.030  &  -4.70E-04\,$\pm$\,0.00631\,$\pm$\,0.00019  &  0.00022  \\
%C2 & ~21.62\,$\pm$\,1.359  &  -8.35E-04\,$\pm$\,0.00590\,$\pm$\,0.00116  & 0.00025  \\
%C3 & ~25.75\,$\pm$\,1.390  &  ~3.52E-03\,$\pm$\,0.00624\,$\pm$\,0.00037  &  0.00026  \\
%C4 & ~31.48\,$\pm$\,1.633  &  ~6.47E-03\,$\pm$\,0.00745\,$\pm$\,0.00076  &  0.00027  \\
%C5 & ~36.77\,$\pm$\,1.887  &  ~1.52E-02\,$\pm$\,0.00770\,$\pm$\,0.00059  &  0.00027  \\
%C6 & ~44.15\,$\pm$\,2.077  &  -4.28E-03\,$\pm$\,0.00956\,$\pm$\,0.00075  &  0.00024  \\
%C7 & ~52.4\,$\pm$\,2.420  &  ~1.08E-02\,$\pm$\,0.01236\,$\pm$\,0.00115  &  0.00024  \\
%C8 & ~62.85\,$\pm$\,2.859  &  ~3.88E-02\,$\pm$\,0.01848\,$\pm$\,0.00150  &  0.00023  \\
%C9 & ~75.97\,$\pm$\,3.435  &  ~5.37E-02\,$\pm$\,0.02688\,$\pm$\,0.00187  &  0.00023  \\
%C10 & ~90.11\,$\pm$\,3.844  &  -9.99E-02\,$\pm$\,0.04329\,$\pm$\,0.00631  &  0.00023  \\
C1 & ~19.81\,$\pm$\,0.81  &  0.0058\,$\pm$\,0.0085\,$\pm$\,0.0002  &  0.00026  \\
C2 & ~22.01\,$\pm$\,1.12  &  -0.0006\,$\pm$\,0.0066\,$\pm$\,0.0006  & 0.00063  \\
C3 & ~26.16\,$\pm$\,1.08  &  -0.0043\,$\pm$\,0.0062\,$\pm$\,0.0004  &  0.00074  \\
C4 & ~32.70\,$\pm$\,1.19  &  0.0049\,$\pm$\,0.0065\,$\pm$\,0.0006  &  0.00078  \\
C5 & ~38.71\,$\pm$\,1.34  &  0.0046\,$\pm$\,0.0077\,$\pm$\,0.0007  &  0.00081  \\
C6 & ~46.01\,$\pm$\,1.36  &  0.0155\,$\pm$\,0.0079\,$\pm$\,0.0006  &  0.00052  \\
C7 & ~53.89\,$\pm$\,1.78  &  -0.0045\,$\pm$\,0.0098\,$\pm$\,0.0012  &  0.00053  \\
C8 & ~64.75\,$\pm$\,1.74  &  0.0104\,$\pm$\,0.0127\,$\pm$\,0.0014  &  0.00041  \\
C9 & ~77.15\,$\pm$\,2.27  &  0.0346\,$\pm$\,0.0192\,$\pm$\,0.0019  &  0.00041  \\
C10 & ~91.14\,$\pm$\,2.58  &  0.0593\,$\pm$\,0.0294\,$\pm$\,0.0073  &  0.00041  \\
\hline
\multicolumn{4}{c}{Topology D:~~~  Forward-Backward Dijets}\\
\hline
%D1 & ~19.8\,$\pm$\,2.231  &  ~3.38E-03\,$\pm$\,0.00629\,$\pm$\,0.00022  &  0.00040  \\
%D2 & ~21.85\,$\pm$\,1.258  &  ~5.56E-03\,$\pm$\,0.00489\,$\pm$\,0.00094  & 0.00022  \\
%D3 & ~25.91\,$\pm$\,1.267  &  -2.35E-03\,$\pm$\,0.00486\,$\pm$\,0.00032  &  0.00029  \\
%D4 & ~30.53\,$\pm$\,1.550  &  -8.00E-04\,$\pm$\,0.00574\,$\pm$\,0.00041  &  0.00025  \\
%D5 & ~36.98\,$\pm$\,1.805  &  -2.36E-03\,$\pm$\,0.00583\,$\pm$\,0.00050  &  0.00026  \\
%D6 & ~43.64\,$\pm$\,2.032  &  ~6.42E-03\,$\pm$\,0.00689\,$\pm$\,0.00094  &  0.00024  \\
%D7 & ~52.62\,$\pm$\,2.417  &  ~5.53E-03\,$\pm$\,0.00850\,$\pm$\,0.00065  &  0.00024  \\
%D8 & ~62.4\,$\pm$\,2.871  &  ~8.05E-03\,$\pm$\,0.01208\,$\pm$\,0.00091  &  0.00023  \\
%D9 & ~75.07\,$\pm$\,3.255  &  ~2.81E-02\,$\pm$\,0.01722\,$\pm$\,0.00208  &  0.00023  \\
%D10 & ~88.83\,$\pm$\,3.696  &  -5.63E-03\,$\pm$\,0.02717\,$\pm$\,0.00150  &  0.00023  \\
%D11 & 106.22\,$\pm$\,4.437  &  -2.24E-02\,$\pm$\,0.04607\,$\pm$\,0.00285  &  0.00023  \\
D1 & ~20.54\,$\pm$\,2.14  &  0.0054\,$\pm$\,0.0161\,$\pm$\,0.0002  &  0.00131  \\
D2 & ~22.09\,$\pm$\,0.95  &  0.0042\,$\pm$\,0.0051\,$\pm$\,0.0002  & 0.00022  \\
D3 & ~26.31\,$\pm$\,0.98  &  0.0051\,$\pm$\,0.0050\,$\pm$\,0.0003  &  0.00072  \\
D4 & ~31.72\,$\pm$\,1.22  &  -0.0031\,$\pm$\,0.0059\,$\pm$\,0.0003  &  0.00052  \\
D5 & ~38.48\,$\pm$\,1.31  &  -0.0018\,$\pm$\,0.0060\,$\pm$\,0.0004  &  0.00054  \\
D6 & ~45.21\,$\pm$\,1.55  &  -0.0040\,$\pm$\,0.0070\,$\pm$\,0.0005  &  0.00044  \\
D7 & ~54.15\,$\pm$\,1.69  &  0.0034\,$\pm$\,0.0087\,$\pm$\,0.0008  &  0.00044  \\
D8 & ~64.30\,$\pm$\,1.92  &  0.0050\,$\pm$\,0.0123\,$\pm$\,0.0011  &  0.00036  \\
D9 & ~76.66\,$\pm$\,2.15  &  0.0058\,$\pm$\,0.0178\,$\pm$\,0.0024  &  0.00036  \\
D10 & ~89.57\,$\pm$\,2.62  &  0.0291\,$\pm$\,0.0296\,$\pm$\,0.0014  &  0.00036  \\
D11 & 106.86\,$\pm$\,2.97  &  -0.0055\,$\pm$\,0.0461\,$\pm$\,0.0028  &  0.00029  \\
\hline\hline
\end{tabular}
\end{table*}

\section{Results and impact}\label{sect:Results}

\begin{figure}[tbh]
\includegraphics*[bb=20 10 510 360,width=\columnwidth]{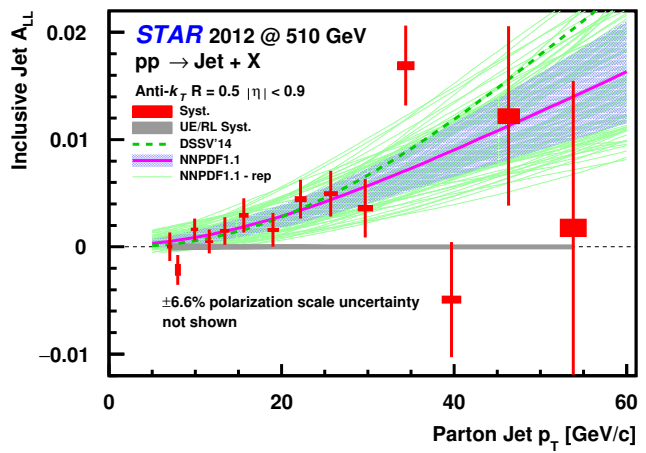}
\caption{\label{fig:inclusiveJet1} $A_{LL}$ as a function of parton jet $p_T$ for inclusive jets with $|\eta|<0.9$ in $\sqrt{s}=510$ GeV $pp$ collisions. The bars show statistical errors, while the size of the boxes show the point-to-point systematic uncertainties on $A_{LL}$ (vertical) and $p_T$ (horizontal).  The gray band on the horizontal axis represents the combined relative luminosity and underlying event uncertainties, which are common to all the points.  The results are compared to predictions from DSSV14 \cite{DSSV14} and NNPDFpol1.1 \cite{NNPDFpol}, including the solid blue uncertainty band for the latter. The green curves are predictions from the 100 equally probable NNPDFpol1.1 replicas.}
\end{figure}

%\begin{figure}[tbh]
%\includegraphics*[bb=20 10 510 360,width=\columnwidth]{figures/run12_asym_final_i_v2.eps}
%\caption{\label{fig:inclusiveJet1} $A_{LL}$ as a function of parton jet $p_T$ for inclusive jets with $|\eta|<0.9$ in $\sqrt{s}=510$ GeV $pp$ collisions. The bars show statistical errors, while the size of the boxes show the point-to-point systematic uncertainties on $A_{LL}$ (vertical) and $p_T$ (horizontal).  The gray band on the horizontal axis represents the combined relative luminosity and underlying event uncertainties, which are common to all the points.  The results are compared to predictions from DSSV14 \cite{DSSV14} and NNPDFpol1.1 %\cite{NNPDFpol}, including the uncertainty band for the latter. T}
%\end{figure}

%\begin{figure}[tbh]
%\includegraphics*[bb=20 10 510 360,width=\columnwidth]{figures/run12_asym_final_nlo_i_v2.eps}
%\caption{\label{fig:inclusiveJet2} $A_{LL}$ as a function of parton jet $p_T$ for inclusive jets with $|\eta|<0.9$ in $\sqrt{s}=510$ GeV $pp$ collisions, compared to predictions from 100 equally probable NNPDFpol1.1 replicas \cite{NNPDFpol}.
%}
%\end{figure}

The inclusive jet and dijet $A_{LL}$ are presented as functions of the fully corrected parton-level jet $p_T$ and dijet $M_{inv}$ in Tables \ref{tab:jet_ALL} and \ref{tab:dijet_ALL}. Figure \ref{fig:inclusiveJet1} shows the inclusive jet asymmetries and systematic uncertainties compared to the theoretical predictions from the DSSV14 \cite{DSSV14} and NNPDFpol1.1 \cite{NNPDFpol} global analyses. The red lines are the statistical errors while the size of the red boxes represent the uncorrelated systematic uncertainties on $A_{LL}$ (vertical) and parton jet $p_T$ (horizontal). The correlated errors, which include the underlying event systematic uncertainty on $A_{LL}$ combined in quadrature with the relative luminosity systematic uncertainty, are plotted as a gray band on the horizontal axis.

\begin{figure}[tbh]
\includegraphics*[bb=20 10 510 360,width=\columnwidth]{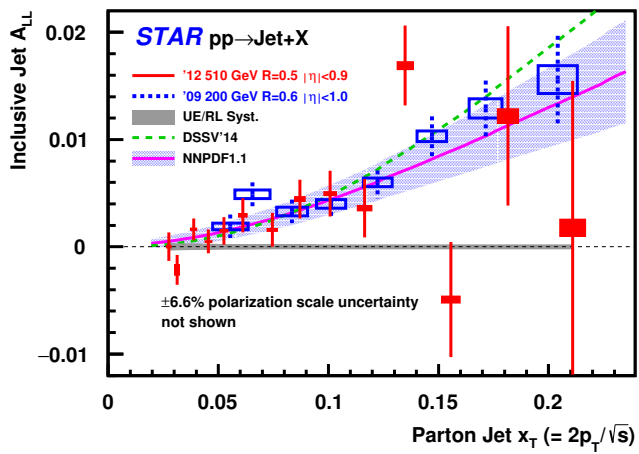}
\caption{\label{fig:inclusiveJet200} $A_{LL}$ as a function of $x_T$ for inclusive jets in $\sqrt{s}=510$ GeV $pp$ collisions (red solid lines), compared to previous measurements of $A_{LL}$ at $\sqrt{s}=200$ GeV (blue dotted lines)~\cite{run9aLL2015}.  The size of the boxes show the systematic uncertainties.  Predictions from DSSV14 \cite{DSSV14} and NNPDFpol1.1 \cite{NNPDFpol}, including the solid blue uncertainty band for the latter, are shown for $\sqrt{s}=510$~GeV.  Predictions for $\sqrt{s}=200$~GeV are similar.
}
\end{figure}

\begin{figure*}[tbh]
\includegraphics*[height=14cm,keepaspectratio]
{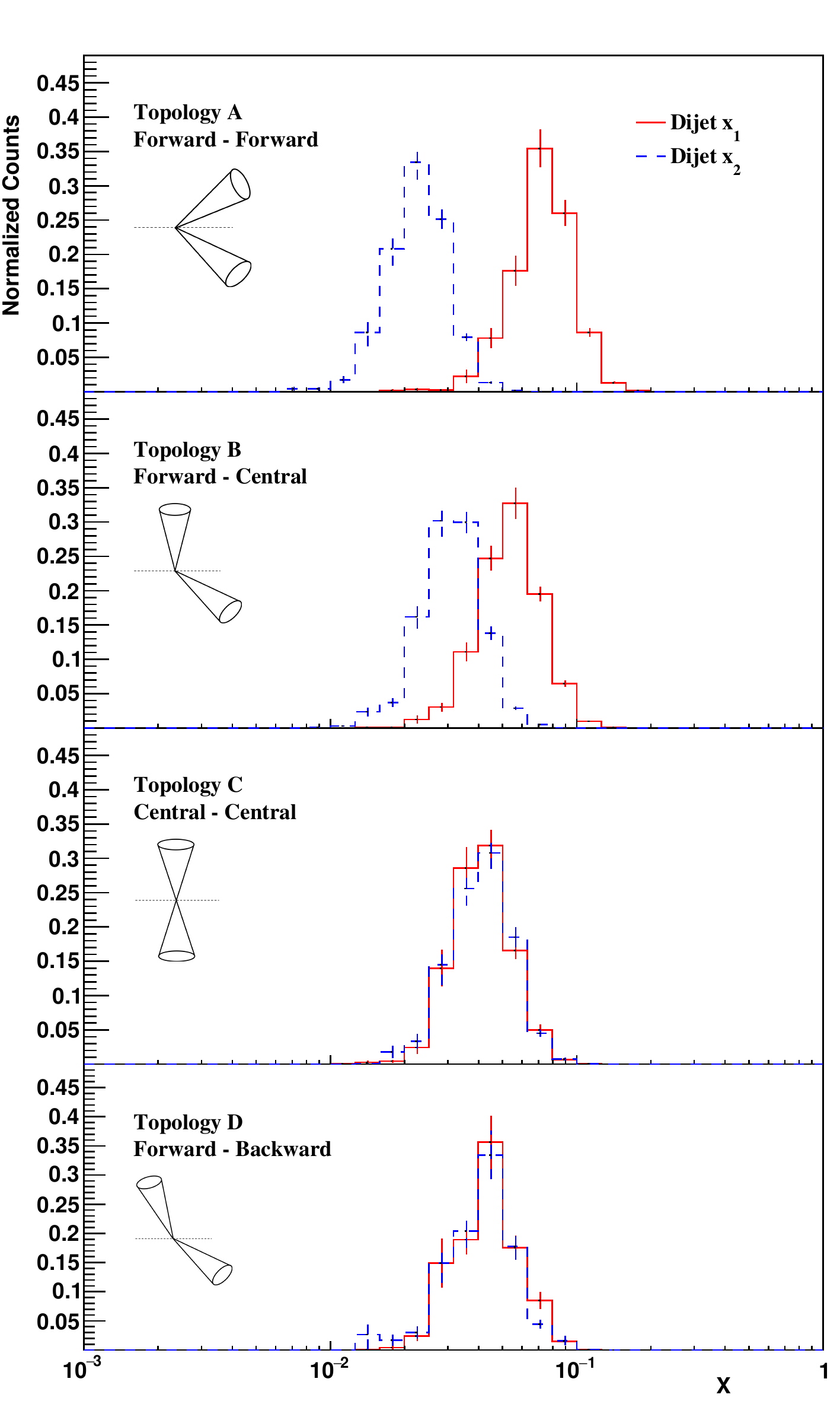}
\includegraphics*[height=14cm, keepaspectratio]
{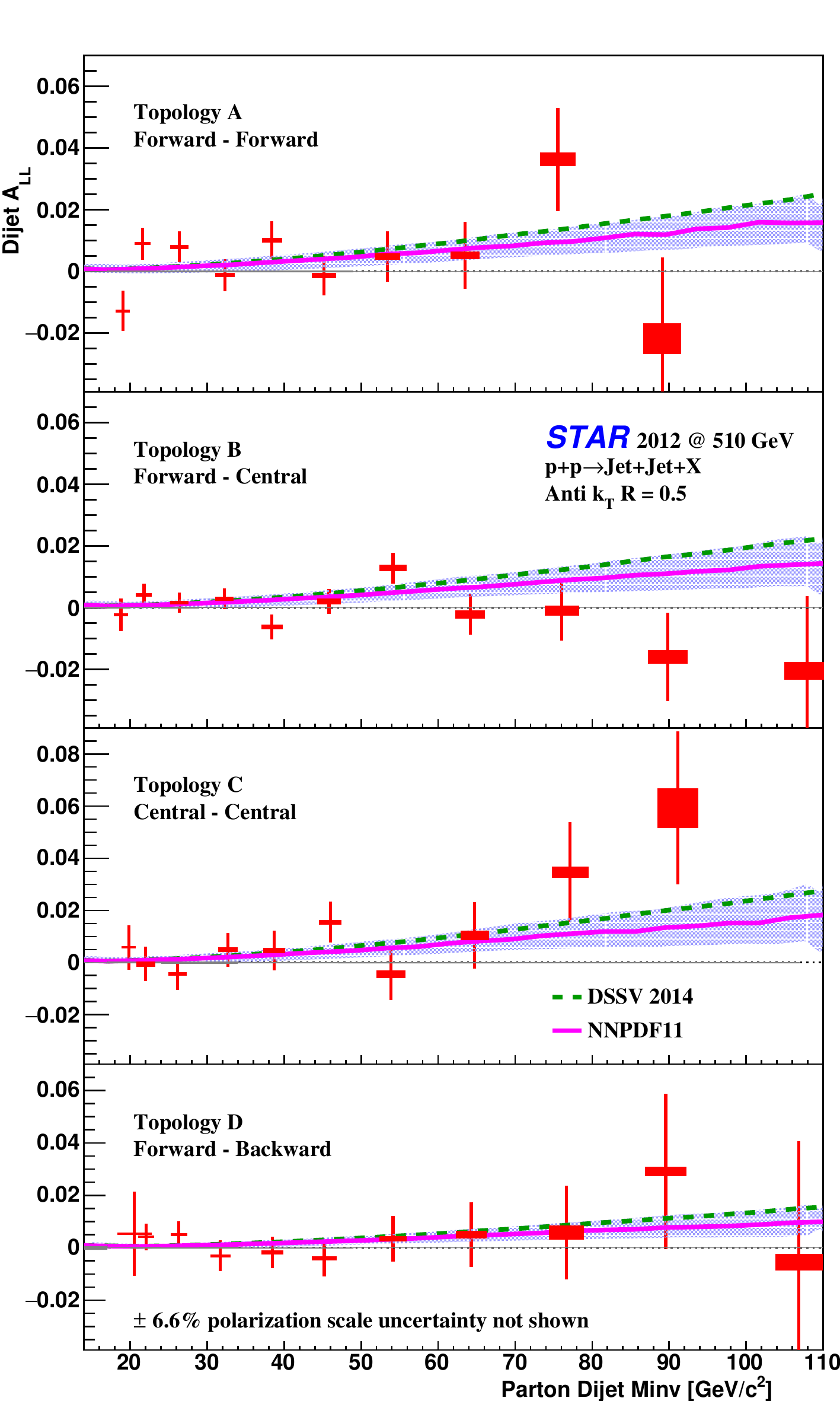}
\caption{\label{fig:DijetXALL} Left: Leading order extractions from the embedding sample of the $x_1$ (red) and $x_2$ (blue) distributions in dijet events with $M_{inv}=17-20$ GeV/$c^2$ for topological bins A-D, which are defined in detail in Table \ref{table:topological}.  The small figures illustrate the approximate orientations of the two jets relative to the beam line. Right: Dijet $A_{LL}$ for the same topological bins in $\sqrt{s} = 510$~GeV $pp$ collisions.  The results are compared to theoretical predictions from DSSV14 and NNPDFpol1.1.} 
\end{figure*}

The theory curves were generated by utilizing the polarized PDFs in the NLO jet production code of Mukherjee and Vogelsang \cite{nlojet2012}. Both theory curves, which include gluon polarization data from RHIC only for energies up to $\sqrt{s}=200$ GeV, show very good agreement with the measured asymmetries. 
%These data do not have enough statistical power at high $p_T$ to discriminate between the best fit predictions from DSSV14 (blue dashed line) and NNPDFpol1.1 (magenta solid line). 
The solid blue region represents the nominal one-sigma error band for NNPDFpol1.1. This uncertainty corresponds to the root-mean-square of the distribution of 100 equally probable replica predictions represented by the green lines. This figure clearly demonstrates the ability of these data to constrain the existing polarized PDF uncertainties, which are driven by uncertainties in the gluon helicity distribution.

In the region of kinematic overlap, the $\sqrt{s}=510$ GeV inclusive jet asymmetries are statistically consistent with the previous measurements at $\sqrt{s} =200$ GeV \cite{run9aLL2015}. This is demonstrated in Fig.\@ \ref{fig:inclusiveJet200}, which compares $A_{LL}$ as a function of $x_T = 2p_T/\sqrt{s}$ for the two energies.  The extended kinematic reach of the 510 GeV asymmetries to lower $x_T$ values is also seen in Fig.\@ \ref{fig:inclusiveJet200}.  The jet $x_T$ is correlated with the initial partonic longitudinal momentum fraction $x$, translating into a sensitivity to lower $x$ partons as well. Figure \ref{fig:inclusiveJetX} shows that the new inclusive jet results extend the sensitivity to gluon polarization down to $x \simeq 0.015$. 

The broad $x_g$ distributions in Fig.\@ \ref{fig:inclusiveJetX} show the wide range of $x$ sampled by each inclusive jet $p_T$ bin. In contrast, the dijets permit full reconstruction of the initial $x_1$ and $x_2$ at leading order. The left-hand side panels in Fig.\@ \ref{fig:DijetXALL} show the leading order extractions, from the embedded simulation sample, of the $x_1$ and $x_2$ distributions for a single dijet bin $M_{inv}$ = $17-20$ GeV/$c^2$. The $x_{1,2}$ values are calculated according to the leading order equations discussed in the introduction using the fully corrected jet $p_T$. The difference between $x_1$ and $x_2$ is largest for topological bin A (top) and decreases until they are identical in topological bins C and D. For all dijet bins, the widths of the $x_1$ and $x_2$ distributions are much narrower than those for the inclusive jet bins, providing a finer resolution on the $x$ dependence of the extracted $\Delta{g(x)}$.

The right-hand side panels in Fig.\@ \ref{fig:DijetXALL} show the dijet $A_{LL}$ as a function of the fully corrected parton-level $M_{inv}$ for the same topological-bin ordering as the left-hand side. The statistical errors are shown by red bars, while the vertical and horizontal width of the red boxes represent the uncorrelated systematic errors on the asymmetries and mass scale, respectively. The correlated errors, which include the underlying event systematic uncertainty on $A_{LL}$ combined in quadrature with the relative luminosity systematic uncertainty, are plotted as a gray band on the horizontal axis. The dijet asymmetries are compared to the same NLO theoretical predictions as in the inclusive case, DSSV14 \cite{DSSV14} and NNPDFpol1.1 \cite{NNPDFpol}. The solid blue bands represent the PDF uncertainties for the NNPDFpol1.1 curves.  Scale uncertainties were also calculated for NNPDFpol1.1 and found to be negligible in comparison to the PDF uncertainties.  The predicted asymmetries are larger for the central-central topology than for the forward-backward combination, even though the $x$ ranges sampled are very similar, because the smaller $\Delta\eta$ between the jets maximizes $\hat{a}_{LL}(\cos(\theta^*))$. Generally the data show good agreement with the theoretical predictions, although with reduced statistical precision compared to the inclusive channel. There are regions, for example at low(high) $M_{inv}$ in bin A(B) where the data will impact future global analyses of the polarized parton distribution, demonstrating the complementarity of the dijet and inclusive jet observables.

\section{Conclusion \label{sect:conclusions}}

We have presented measurements of the longitudinal double-spin asymmetry $A_{LL}$ for inclusive jet and dijet production at midrapidity in $pp$ collisions at $\sqrt{s} = 510$~GeV, based on data that were recorded by the STAR Collaboration during the 2012 RHIC running period.  The results are sensitive to the gluon polarization over the momentum fraction range from $x \approx 0.015$ to $x \approx 0.2$. The inclusive jet results will provide important new constraints on the magnitude of the gluon polarization and the dijet results will provide important new constraints on the shape of $\Delta g(x)$ when they are included in future global analyses of the polarized PDFs, especially in the region $x < 0.05$ that has been unconstrained by input data in previous global analyses.

\appendix*

\section{Correlation matrix}

The inclusive jet and dijet $A_{LL}$ results presented here have two systematic uncertainties that are common to all the data points.  The relative luminosity uncertainty represents a common offset of the $A_{LL}$ = 0 axis by $\pm\,2.2 \times 10^{-4}$.  An uncertainty of $\pm$\,6.6\% in the product of the beam polarizations represents an overall scale uncertainty. In addition, there are point-to-point statistical and systematic correlations, as discussed in Sect.\@ \ref{sect:correlations}.  The correlation matrix that quantifies these additional point-to-point effects is given in Tables \ref{tab:corrIVsI}--\ref{tab:corrDVsD}.  The entries that are not shown can be obtained by transposition.

\begin{table*}[h]
\caption{The correlation matrix for the point-to-point uncertainties in the inclusive jet measurements.  At low $p_T$, the dominant effects arise from correlated systematic uncertainties, whereas at high $p_T$, the dominant effects arise from the statistical correlations when two jets in the same event satisfy all the inclusive jet cuts.  The relative luminosity and beam polarization uncertainties, which are common to all the data points, are not included.}
\label{tab:corrIVsI}
\begin{tabular}{c@{~~~}c@{~~}c@{~~}c@{~~}c@{~~}c@{~~}c@{~~}c@{~~}c@{~~}c@{~~}c@{~~}c@{~~}c@{~~}c@{~~}c}
\hline 
 \hline 
Bin	& I1 & I2 & I3 & I4 & I5 & I6 & I7 & I8 & I9 & I10 & I11 & I12 & I13 & I14 \\ 
\hline 
I1	&  1 &  0.065 &  0.057 &  0.044 &  0.036 &  0.026 &  0.024 &  0.024 &  0.016 &  0.013 &  0.011 &  0.006 &  0.007 &  0.005 \\  
I2	&   &  1 &  0.056 &  0.045 &  0.039 &  0.030 &  0.031 &  0.036 &  0.023 &  0.023 &  0.020 &  0.010 &  0.016 &  0.011 \\  
I3	&   &   &  1 &  0.046 &  0.039 &  0.030 &  0.027 &  0.024 &  0.016 &  0.013 &  0.009 &  0.005 &  0.006 &  0.004 \\  
I4	&   &   &   &  1 &  0.035 &  0.028 &  0.025 &  0.022 &  0.014 &  0.011 &  0.008 &  0.005 &  0.005 &  0.003 \\  
I5	&   &   &   &   &  1 &  0.029 &  0.026 &  0.023 &  0.015 &  0.012 &  0.009 &  0.005 &  0.005 &  0.003 \\  
I6	&   &   &   &   &   &  1 &  0.025 &  0.023 &  0.016 &  0.012 &  0.009 &  0.005 &  0.005 &  0.003 \\  
I7	&   &   &   &  &   &   &  1 &  0.028 &  0.022 &  0.019 &  0.014 &  0.008 &  0.008 &  0.005 \\  
I8	&   &   &   &   &   &  &   &  1 &  0.029 &  0.027 &  0.022 &  0.014 &  0.012 &  0.008 \\  
I9	&   &   &   &   &   &   &   &   &  1 &  0.030 &  0.026 &  0.018 &  0.014 &  0.008 \\  
I10	&  &   &   &  &   &   &   &   &   &  1 &  0.035 &  0.028 &  0.022 &  0.013 \\  
I11	&   &  &   &   &  &   &   &   &  &   &  1 &  0.037 &  0.031 &  0.020 \\  
I12	&   &   &   &   &   &   &   &   &   &   &   &  1 &  0.041 &  0.030 \\  
I13	&   &   &   &   &   &   &  &   &   &   &   &   &  1 &  0.044 \\  
I14	&   &   &   &   &   &   &   &   &   &   &   &   &   &  1 \\  
\hline 
 \hline 
\end{tabular}
\end{table*}

\begin{table*}[h]
\caption{The correlation matrix for the point-to-point uncertainties coupling the inclusive jet measurements with the forward-forward dijet measurements (Topology A).  The relative luminosity and beam polarization uncertainties, which are common to all the data points, are not included.}
\label{tab:corrIVsA}
\begin{tabular}{c@{~~~}c@{~~}c@{~~}c@{~~}c@{~~}c@{~~}c@{~~}c@{~~}c@{~~}c@{~~}c}
\hline 
 \hline 
Bin	& A1 & A2 & A3 & A4 & A5 & A6 & A7 & A8 & A9 & A10 \\ 
\hline 
I1	&  0.020 &  0.011 &  0.009 &  0.007 &  0.007 &  0.009 &  0.010 &  0.009 &  0.011 &  0.017 \\  
I2	&  0.027 &  0.025 &  0.023 &  0.018 &  0.018 &  0.023 &  0.027 &  0.023 &  0.029 &  0.044 \\  
I3	&  0.066 &  0.054 &  0.029 &  0.013 &  0.006 &  0.006 &  0.007 &  0.006 &  0.008 &  0.012 \\  
I4	&  0.033 &  0.062 &  0.056 &  0.025 &  0.009 &  0.006 &  0.006 &  0.006 &  0.007 &  0.010 \\  
I5	&  0.006 &  0.053 &  0.064 &  0.053 &  0.020 &  0.010 &  0.007 &  0.006 &  0.007 &  0.011 \\  
I6	&  0.002 &  0.023 &  0.058 &  0.064 &  0.042 &  0.018 &  0.008 &  0.005 &  0.007 &  0.010 \\  
I7	&  0.002 &  0.005 &  0.047 &  0.072 &  0.078 &  0.059 &  0.024 &  0.009 &  0.008 &  0.012 \\  
I8	&  0.003 &  0.005 &  0.026 &  0.059 &  0.073 &  0.096 &  0.059 &  0.023 &  0.014 &  0.019 \\  
I9	&  0.002 &  0.003 &  0.008 &  0.034 &  0.052 &  0.081 &  0.103 &  0.052 &  0.018 &  0.012 \\  
I10	&  0.002 &  0.003 &  0.005 &  0.016 &  0.031 &  0.056 &  0.086 &  0.116 &  0.048 &  0.021 \\  
I11	&  0.002 &  0.003 &  0.005 &  0.007 &  0.015 &  0.031 &  0.057 &  0.093 &  0.124 &  0.047 \\  
I12	&  0.001 &  0.002 &  0.003 &  0.003 &  0.005 &  0.012 &  0.026 &  0.057 &  0.096 &  0.131 \\  
I13	&  0.002 &  0.003 &  0.005 &  0.004 &  0.005 &  0.008 &  0.014 &  0.029 &  0.063 &  0.116 \\  
I14	&  0.001 &  0.002 &  0.003 &  0.003 &  0.003 &  0.004 &  0.006 &  0.011 &  0.026 &  0.064 \\  
\hline 
 \hline 
\end{tabular}
\end{table*}

\begin{table*}[h]
\caption{The correlation matrix for the point-to-point uncertainties coupling the inclusive jet measurements with the forward-central dijet measurements (Topology B).  The relative luminosity and beam polarization uncertainties, which are common to all the data points, are not included.}
\label{tab:corrIVsB}
\begin{tabular}{c@{~~~}c@{~~}c@{~~}c@{~~}c@{~~}c@{~~}c@{~~}c@{~~}c@{~~}c@{~~}c@{~~}c}
\hline 
 \hline 
Bin	& B1 & B2 & B3 & B4 & B5 & B6 & B7 & B8 & B9 & B10 & B11 \\ 
\hline 
I1	&  0.028 &  0.020 &  0.013 &  0.012 &  0.011 &  0.015 &  0.015 &  0.016 &  0.013 &  0.013 &  0.010 \\  
I2	&  0.033 &  0.043 &  0.035 &  0.032 &  0.028 &  0.038 &  0.038 &  0.041 &  0.034 &  0.034 &  0.027 \\  
I3	&  0.086 &  0.085 &  0.051 &  0.024 &  0.011 &  0.011 &  0.010 &  0.011 &  0.009 &  0.009 &  0.007 \\  
I4	&  0.032 &  0.090 &  0.087 &  0.046 &  0.018 &  0.012 &  0.009 &  0.010 &  0.008 &  0.008 &  0.006 \\  
I5	&  0.005 &  0.063 &  0.094 &  0.089 &  0.037 &  0.019 &  0.011 &  0.011 &  0.009 &  0.009 &  0.007 \\  
I6	&  0.002 &  0.022 &  0.082 &  0.094 &  0.078 &  0.035 &  0.014 &  0.010 &  0.008 &  0.008 &  0.006 \\  
I7	&  0.003 &  0.006 &  0.055 &  0.101 &  0.109 &  0.103 &  0.043 &  0.019 &  0.010 &  0.010 &  0.008 \\  
I8	&  0.004 &  0.008 &  0.027 &  0.080 &  0.103 &  0.138 &  0.105 &  0.047 &  0.020 &  0.015 &  0.012 \\  
I9	&  0.002 &  0.005 &  0.008 &  0.043 &  0.071 &  0.114 &  0.147 &  0.101 &  0.033 &  0.012 &  0.007 \\  
I10	&  0.003 &  0.006 &  0.007 &  0.020 &  0.041 &  0.078 &  0.119 &  0.171 &  0.094 &  0.031 &  0.010 \\  
I11	&  0.003 &  0.005 &  0.007 &  0.010 &  0.020 &  0.043 &  0.077 &  0.131 &  0.185 &  0.089 &  0.023 \\  
I12	&  0.001 &  0.003 &  0.003 &  0.004 &  0.006 &  0.017 &  0.035 &  0.076 &  0.134 &  0.200 &  0.072 \\  
I13	&  0.002 &  0.005 &  0.006 &  0.008 &  0.007 &  0.012 &  0.019 &  0.040 &  0.081 &  0.156 &  0.208 \\  
I14	&  0.002 &  0.003 &  0.004 &  0.005 &  0.005 &  0.006 &  0.008 &  0.016 &  0.035 &  0.086 &  0.179 \\  
\hline 
 \hline 
\end{tabular}
\end{table*}

\begin{table*}[h]
\caption{The correlation matrix for the point-to-point uncertainties coupling the inclusive jet measurements with the central-central dijet measurements (Topology C).  The relative luminosity and beam polarization uncertainties, which are common to all the data points, are not included.}
\begin{tabular}{c@{~~~}c@{~~}c@{~~}c@{~~}c@{~~}c@{~~}c@{~~}c@{~~}c@{~~}c@{~~}c}
\hline 
 \hline 
Bin	& C1 & C2 & C3 & C4 & C5 & C6 & C7 & C8 & C9 & C10 \\ 
\hline 
I1	&  0.016 &  0.014 &  0.008 &  0.008 &  0.008 &  0.007 &  0.011 &  0.011 &  0.009 &  0.023 \\  
I2	&  0.021 &  0.034 &  0.020 &  0.022 &  0.022 &  0.018 &  0.029 &  0.027 &  0.023 &  0.059 \\  
I3	&  0.051 &  0.045 &  0.024 &  0.012 &  0.007 &  0.005 &  0.008 &  0.007 &  0.006 &  0.016 \\  
I4	&  0.026 &  0.052 &  0.044 &  0.022 &  0.009 &  0.005 &  0.007 &  0.007 &  0.006 &  0.014 \\  
I5	&  0.004 &  0.046 &  0.052 &  0.045 &  0.018 &  0.008 &  0.008 &  0.007 &  0.006 &  0.015 \\  
I6	&  0.001 &  0.021 &  0.047 &  0.055 &  0.038 &  0.015 &  0.008 &  0.006 &  0.005 &  0.013 \\  
I7	&  0.002 &  0.008 &  0.036 &  0.057 &  0.062 &  0.046 &  0.021 &  0.010 &  0.007 &  0.017 \\  
I8	&  0.002 &  0.010 &  0.020 &  0.048 &  0.060 &  0.077 &  0.051 &  0.023 &  0.012 &  0.025 \\  
I9	&  0.001 &  0.006 &  0.006 &  0.028 &  0.044 &  0.065 &  0.086 &  0.045 &  0.015 &  0.016 \\  
I10	&  0.002 &  0.007 &  0.005 &  0.015 &  0.027 &  0.046 &  0.073 &  0.096 &  0.041 &  0.025 \\  
I11	&  0.002 &  0.006 &  0.004 &  0.007 &  0.015 &  0.025 &  0.049 &  0.080 &  0.102 &  0.047 \\  
I12	&  0.001 &  0.003 &  0.002 &  0.003 &  0.005 &  0.010 &  0.024 &  0.051 &  0.083 &  0.112 \\  
I13	&  0.001 &  0.006 &  0.004 &  0.005 &  0.006 &  0.006 &  0.014 &  0.027 &  0.055 &  0.108 \\  
I14	&  0.001 &  0.004 &  0.003 &  0.004 &  0.004 &  0.003 &  0.006 &  0.012 &  0.024 &  0.064 \\  
\hline 
 \hline 
\end{tabular}
\label{tab:corrIVsC}
\end{table*}

\begin{table*}[h]
\caption{The correlation matrix for the point-to-point uncertainties coupling the inclusive jet measurements with the forward-backward dijet measurements (Topology D).  The relative luminosity and beam polarization uncertainties, which are common to all the data points, are not included.}
\label{tab:corrIVsD}
\begin{tabular}{c@{~~~}c@{~~}c@{~~}c@{~~}c@{~~}c@{~~}c@{~~}c@{~~}c@{~~}c@{~~}c@{~~}c}
\hline 
 \hline 
Bin	& D1 & D2 & D3 & D4 & D5 & D6 & D7 & D8 & D9 & D10 & D11 \\ 
\hline 
I1	&  0.011 &  0.018 &  0.011 &  0.006 &  0.007 &  0.007 &  0.008 &  0.009 &  0.012 &  0.005 &  0.006 \\  
I2	&  0.011 &  0.029 &  0.026 &  0.017 &  0.017 &  0.019 &  0.022 &  0.022 &  0.032 &  0.012 &  0.015 \\  
I3	&  0.030 &  0.060 &  0.047 &  0.023 &  0.010 &  0.006 &  0.006 &  0.006 &  0.009 &  0.003 &  0.004 \\  
I4	&  0.004 &  0.042 &  0.061 &  0.043 &  0.019 &  0.009 &  0.006 &  0.005 &  0.008 &  0.003 &  0.004 \\  
I5	&  0.001 &  0.016 &  0.056 &  0.061 &  0.038 &  0.018 &  0.008 &  0.006 &  0.008 &  0.003 &  0.004 \\  
I6	&  0.001 &  0.004 &  0.034 &  0.057 &  0.057 &  0.037 &  0.014 &  0.006 &  0.007 &  0.003 &  0.003 \\  
I7	&  0.001 &  0.003 &  0.015 &  0.054 &  0.065 &  0.076 &  0.045 &  0.018 &  0.011 &  0.004 &  0.004 \\  
I8	&  0.001 &  0.005 &  0.008 &  0.034 &  0.056 &  0.078 &  0.085 &  0.047 &  0.023 &  0.006 &  0.007 \\  
I9	&  0.001 &  0.003 &  0.004 &  0.013 &  0.035 &  0.059 &  0.080 &  0.089 &  0.042 &  0.011 &  0.005 \\  
I10	&  0.001 &  0.003 &  0.004 &  0.005 &  0.018 &  0.036 &  0.060 &  0.091 &  0.096 &  0.036 &  0.010 \\  
I11	&  0.001 &  0.003 &  0.004 &  0.004 &  0.008 &  0.019 &  0.035 &  0.064 &  0.101 &  0.098 &  0.030 \\  
I12	&  0.000 &  0.002 &  0.002 &  0.002 &  0.003 &  0.006 &  0.015 &  0.032 &  0.064 &  0.109 &  0.094 \\  
I13	&  0.001 &  0.003 &  0.004 &  0.003 &  0.004 &  0.005 &  0.009 &  0.016 &  0.036 &  0.068 &  0.127 \\  
I14	&  0.001 &  0.002 &  0.002 &  0.002 &  0.003 &  0.003 &  0.004 &  0.006 &  0.016 &  0.033 &  0.076 \\  
\hline 
 \hline 
\end{tabular}
%\end{table*}

\vspace{0.4in}

%\begin{table*}[h]
\caption{The correlation matrix for the point-to-point uncertainties in the forward-forward dijet measurements (Topology A).  The relative luminosity and beam polarization uncertainties, which are common to all the data points, are not included.}
\label{tab:corrAVsA}
\begin{tabular}{c@{~~~}c@{~~}c@{~~}c@{~~}c@{~~}c@{~~}c@{~~}c@{~~}c@{~~}c@{~~}c}
\hline 
 \hline 
Bin	& A1 & A2 & A3 & A4 & A5 & A6 & A7 & A8 & A9 & A10 \\ 
\hline 
A1	&  1 &  0.003 &  0.004 &  0.004 &  0.004 &  0.004 &  0.004 &  0.003 &  0.004 &  0.006 \\  
A2	&   &  1 &  0.015 &  0.013 &  0.011 &  0.011 &  0.010 &  0.007 &  0.007 &  0.009 \\  
A3	&   &   &  1 &  0.016 &  0.014 &  0.014 &  0.013 &  0.010 &  0.011 &  0.014 \\  
A4	&   &   &   &  1 &  0.012 &  0.013 &  0.012 &  0.009 &  0.010 &  0.013 \\  
A5	&   &   &   &   &  1 &  0.011 &  0.011 &  0.009 &  0.010 &  0.013 \\  
A6	&   &   &   &   &   &  1 &  0.013 &  0.010 &  0.012 &  0.017 \\  
A7	&   &   &   &   &   &   &  1 &  0.012 &  0.014 &  0.020 \\  
A8	&   &   &   &   &   &   &   &  1 &  0.012 &  0.017 \\  
A9	&   &   &   &   &   &   &   &   &  1 &  0.021 \\  
A10	&   &   &   &   &   &   &   &   &   &  1 \\  
\hline 
 \hline 
\end{tabular}
%\end{table*}

\vspace{0.4in}

%\begin{table*}[h]
\caption{The correlation matrix for the point-to-point uncertainties coupling the forward-forward dijet measurements (Topology A) with the forward-central dijet measurements (Topology B).  The relative luminosity and beam polarization uncertainties, which are common to all the data points, are not included.}
\label{tab:corrAVsB}
\begin{tabular}{c@{~~~}c@{~~}c@{~~}c@{~~}c@{~~}c@{~~}c@{~~}c@{~~}c@{~~}c@{~~}c@{~~}c}
\hline 
 \hline 
Bin	& B1 & B2 & B3 & B4 & B5 & B6 & B7 & B8 & B9 & B10 & B11 \\ 
\hline 
A1	&  0.001 &  0.002 &  0.003 &  0.004 &  0.004 &  0.005 &  0.005 &  0.005 &  0.004 &  0.004 &  0.004 \\  
A2	&  0.002 &  0.003 &  0.004 &  0.005 &  0.005 &  0.007 &  0.007 &  0.007 &  0.006 &  0.006 &  0.005 \\  
A3	&  0.003 &  0.006 &  0.007 &  0.009 &  0.008 &  0.011 &  0.011 &  0.012 &  0.010 &  0.010 &  0.008 \\  
A4	&  0.003 &  0.005 &  0.007 &  0.009 &  0.008 &  0.011 &  0.011 &  0.012 &  0.010 &  0.010 &  0.008 \\  
A5	&  0.003 &  0.005 &  0.007 &  0.009 &  0.008 &  0.011 &  0.011 &  0.012 &  0.010 &  0.010 &  0.008 \\  
A6	&  0.004 &  0.007 &  0.009 &  0.011 &  0.010 &  0.015 &  0.014 &  0.016 &  0.013 &  0.013 &  0.010 \\  
A7	&  0.004 &  0.008 &  0.011 &  0.013 &  0.012 &  0.017 &  0.017 &  0.018 &  0.015 &  0.015 &  0.012 \\  
A8	&  0.004 &  0.007 &  0.009 &  0.011 &  0.010 &  0.015 &  0.014 &  0.016 &  0.013 &  0.013 &  0.010 \\  
A9	&  0.004 &  0.009 &  0.012 &  0.014 &  0.013 &  0.019 &  0.018 &  0.020 &  0.017 &  0.016 &  0.013 \\  
A10	&  0.007 &  0.014 &  0.018 &  0.022 &  0.020 &  0.028 &  0.027 &  0.030 &  0.025 &  0.025 &  0.020 \\  
\hline 
 \hline 
\end{tabular}
\end{table*}

\begin{table*}[h]
\caption{The correlation matrix for the point-to-point uncertainties coupling the forward-forward dijet measurements (Topology A) with the central-central dijet measurements (Topology C).  The relative luminosity and beam polarization uncertainties, which are common to all the data points, are not included.}
\label{tab:corrAVsC}
\begin{tabular}{c@{~~~}c@{~~}c@{~~}c@{~~}c@{~~}c@{~~}c@{~~}c@{~~}c@{~~}c@{~~}c}
\hline 
 \hline 
Bin	& C1 & C2 & C3 & C4 & C5 & C6 & C7 & C8 & C9 & C10 \\ 
\hline 
A1	&  0.001 &  0.003 &  0.002 &  0.003 &  0.003 &  0.002 &  0.004 &  0.004 &  0.003 &  0.008 \\  
A2	&  0.001 &  0.004 &  0.003 &  0.004 &  0.004 &  0.003 &  0.005 &  0.005 &  0.004 &  0.011 \\  
A3	&  0.002 &  0.007 &  0.005 &  0.006 &  0.007 &  0.005 &  0.009 &  0.008 &  0.007 &  0.018 \\  
A4	&  0.002 &  0.007 &  0.005 &  0.006 &  0.006 &  0.005 &  0.008 &  0.008 &  0.007 &  0.017 \\  
A5	&  0.002 &  0.007 &  0.005 &  0.006 &  0.006 &  0.005 &  0.008 &  0.008 &  0.007 &  0.017 \\  
A6	&  0.002 &  0.009 &  0.006 &  0.008 &  0.008 &  0.007 &  0.011 &  0.010 &  0.009 &  0.023 \\  
A7	&  0.002 &  0.010 &  0.007 &  0.009 &  0.010 &  0.008 &  0.013 &  0.012 &  0.010 &  0.026 \\  
A8	&  0.002 &  0.009 &  0.006 &  0.008 &  0.008 &  0.007 &  0.011 &  0.011 &  0.009 &  0.023 \\  
A9	&  0.003 &  0.011 &  0.008 &  0.010 &  0.011 &  0.009 &  0.014 &  0.013 &  0.011 &  0.029 \\  
A10	&  0.004 &  0.017 &  0.012 &  0.015 &  0.016 &  0.013 &  0.021 &  0.020 &  0.017 &  0.043 \\  
\hline 
 \hline 
\end{tabular}
%\end{table*}

\vspace{0.4in}

%\begin{table*}[h]
\caption{The correlation matrix for the point-to-point uncertainties coupling the forward-forward dijet measurements (Topology A) with the forward-backward dijet measurements (Topology D).  The relative luminosity and beam polarization uncertainties, which are common to all the data points, are not included.}
\label{tab:corrAVsD}
\begin{tabular}{c@{~~~}c@{~~}c@{~~}c@{~~}c@{~~}c@{~~}c@{~~}c@{~~}c@{~~}c@{~~}c@{~~}c}
\hline 
 \hline 
Bin	& D1 & D2 & D3 & D4 & D5 & D6 & D7 & D8 & D9 & D10 & D11 \\ 
\hline 
A1	&  0.000 &  0.001 &  0.002 &  0.002 &  0.002 &  0.002 &  0.003 &  0.003 &  0.004 &  0.002 &  0.002 \\  
A2	&  0.001 &  0.002 &  0.003 &  0.002 &  0.003 &  0.003 &  0.004 &  0.004 &  0.006 &  0.002 &  0.003 \\  
A3	&  0.001 &  0.003 &  0.004 &  0.004 &  0.005 &  0.006 &  0.006 &  0.007 &  0.010 &  0.004 &  0.004 \\  
A4	&  0.001 &  0.003 &  0.004 &  0.004 &  0.005 &  0.005 &  0.006 &  0.006 &  0.009 &  0.003 &  0.004 \\  
A5	&  0.001 &  0.003 &  0.004 &  0.004 &  0.005 &  0.005 &  0.006 &  0.007 &  0.010 &  0.003 &  0.004 \\  
A6	&  0.001 &  0.004 &  0.005 &  0.005 &  0.006 &  0.007 &  0.008 &  0.008 &  0.012 &  0.004 &  0.006 \\  
A7	&  0.001 &  0.005 &  0.006 &  0.006 &  0.007 &  0.008 &  0.010 &  0.010 &  0.014 &  0.005 &  0.007 \\  
A8	&  0.001 &  0.004 &  0.005 &  0.005 &  0.006 &  0.007 &  0.008 &  0.009 &  0.012 &  0.005 &  0.006 \\  
A9	&  0.002 &  0.006 &  0.007 &  0.006 &  0.008 &  0.009 &  0.010 &  0.011 &  0.016 &  0.006 &  0.007 \\  
A10	&  0.002 &  0.008 &  0.010 &  0.010 &  0.012 &  0.013 &  0.016 &  0.016 &  0.024 &  0.009 &  0.011 \\  
\hline 
 \hline 
\end{tabular}
%\end{table*}

\vspace{0.4in}

%\begin{table*}[h]
\caption{The correlation matrix for the point-to-point uncertainties in the forward-central dijet measurements (Topology B).  The relative luminosity and beam polarization uncertainties, which are common to all the data points, are not included.}
\label{tab:corrBVsB}
\begin{tabular}{c@{~~~}c@{~~}c@{~~}c@{~~}c@{~~}c@{~~}c@{~~}c@{~~}c@{~~}c@{~~}c@{~~}c}
\hline 
 \hline 
Bin	& B1 & B2 & B3 & B4 & B5 & B6 & B7 & B8 & B9 & B10 & B11 \\ 
\hline 
B1	&  1 &  0.013 &  0.010 &  0.011 &  0.008 &  0.010 &  0.008 &  0.008 &  0.006 &  0.006 &  0.005 \\  
B2	&   &  1 &  0.024 &  0.027 &  0.019 &  0.023 &  0.018 &  0.018 &  0.014 &  0.012 &  0.009 \\  
B3	&   &   &  1 &  0.023 &  0.018 &  0.023 &  0.019 &  0.020 &  0.016 &  0.015 &  0.012 \\  
B4	&   &   &   &  1 &  0.021 &  0.026 &  0.023 &  0.024 &  0.019 &  0.018 &  0.014 \\  
B5	&   &   &   &   &  1 &  0.022 &  0.020 &  0.021 &  0.017 &  0.016 &  0.013 \\  
B6	&   &   &   &   &   &  1 &  0.027 &  0.029 &  0.023 &  0.022 &  0.018 \\  
B7	&   &   &   &   &   &   &  1 &  0.027 &  0.022 &  0.022 &  0.017 \\  
B8	&   &   &   &   &   &   &   &  1 &  0.024 &  0.024 &  0.019 \\  
B9	&   &   &   &   &   &   &   &   &  1 &  0.020 &  0.016 \\  
B10	&   &   &   &   &   &   &   &   &   &  1 &  0.015 \\  
B11	&   &   &   &   &   &   &   &   &   &   &  1 \\  
\hline 
 \hline 
\end{tabular}
\end{table*}

\begin{table*}[h]
\caption{The correlation matrix for the point-to-point uncertainties coupling the forward-central dijet measurements (Topology B) with the central-central dijet measurements (Topology C).  The relative luminosity and beam polarization uncertainties, which are common to all the data points, are not included.}
\label{tab:corrBVsC}
\begin{tabular}{c@{~~~}c@{~~}c@{~~}c@{~~}c@{~~}c@{~~}c@{~~}c@{~~}c@{~~}c@{~~}c}
\hline 
 \hline 
Bin	& C1 & C2 & C3 & C4 & C5 & C6 & C7 & C8 & C9 & C10 \\ 
\hline 
B1	&  0.001 &  0.004 &  0.002 &  0.003 &  0.003 &  0.003 &  0.004 &  0.004 &  0.004 &  0.009 \\  
B2	&  0.002 &  0.007 &  0.005 &  0.007 &  0.007 &  0.006 &  0.009 &  0.009 &  0.007 &  0.018 \\  
B3	&  0.002 &  0.009 &  0.006 &  0.009 &  0.009 &  0.007 &  0.012 &  0.011 &  0.009 &  0.024 \\  
B4	&  0.003 &  0.012 &  0.008 &  0.011 &  0.011 &  0.009 &  0.014 &  0.014 &  0.012 &  0.029 \\  
B5	&  0.002 &  0.011 &  0.007 &  0.010 &  0.010 &  0.008 &  0.013 &  0.012 &  0.011 &  0.027 \\  
B6	&  0.003 &  0.015 &  0.010 &  0.014 &  0.014 &  0.011 &  0.018 &  0.017 &  0.015 &  0.038 \\  
B7	&  0.003 &  0.015 &  0.010 &  0.013 &  0.014 &  0.011 &  0.018 &  0.017 &  0.015 &  0.037 \\  
B8	&  0.004 &  0.016 &  0.011 &  0.015 &  0.015 &  0.012 &  0.020 &  0.019 &  0.016 &  0.041 \\  
B9	&  0.003 &  0.013 &  0.009 &  0.012 &  0.013 &  0.010 &  0.017 &  0.016 &  0.014 &  0.034 \\  
B10	&  0.003 &  0.013 &  0.009 &  0.012 &  0.012 &  0.010 &  0.016 &  0.015 &  0.013 &  0.033 \\  
B11	&  0.002 &  0.010 &  0.007 &  0.010 &  0.010 &  0.008 &  0.013 &  0.012 &  0.011 &  0.027 \\  
\hline 
 \hline 
\end{tabular}
%\end{table*}

\vspace{0.4in}

%\begin{table*}[h]
\caption{The correlation matrix for the point-to-point uncertainties coupling the forward-central dijet measurements (Topology B) with the forward-backward dijet measurements (Topology D).  The relative luminosity and beam polarization uncertainties, which are common to all the data points, are not included.}
\label{tab:corrBVsD}
\begin{tabular}{c@{~~~}c@{~~}c@{~~}c@{~~}c@{~~}c@{~~}c@{~~}c@{~~}c@{~~}c@{~~}c@{~~}c}
\hline 
 \hline 
Bin	& D1 & D2 & D3 & D4 & D5 & D6 & D7 & D8 & D9 & D10 & D11 \\ 
\hline 
B1	&  0.001 &  0.002 &  0.002 &  0.002 &  0.003 &  0.003 &  0.003 &  0.003 &  0.005 &  0.002 &  0.002 \\  
B2	&  0.001 &  0.004 &  0.004 &  0.004 &  0.005 &  0.006 &  0.007 &  0.007 &  0.010 &  0.004 &  0.005 \\  
B3	&  0.001 &  0.005 &  0.006 &  0.005 &  0.007 &  0.007 &  0.009 &  0.009 &  0.013 &  0.005 &  0.006 \\  
B4	&  0.002 &  0.006 &  0.007 &  0.007 &  0.008 &  0.009 &  0.011 &  0.011 &  0.016 &  0.006 &  0.007 \\  
B5	&  0.002 &  0.005 &  0.006 &  0.006 &  0.008 &  0.008 &  0.010 &  0.010 &  0.015 &  0.005 &  0.007 \\  
B6	&  0.002 &  0.007 &  0.009 &  0.008 &  0.011 &  0.012 &  0.014 &  0.014 &  0.021 &  0.007 &  0.010 \\  
B7	&  0.002 &  0.007 &  0.009 &  0.008 &  0.010 &  0.012 &  0.014 &  0.014 &  0.020 &  0.007 &  0.009 \\  
B8	&  0.002 &  0.008 &  0.010 &  0.009 &  0.011 &  0.013 &  0.015 &  0.015 &  0.022 &  0.008 &  0.010 \\  
B9	&  0.002 &  0.007 &  0.008 &  0.008 &  0.010 &  0.011 &  0.012 &  0.013 &  0.019 &  0.007 &  0.009 \\  
B10	&  0.002 &  0.006 &  0.008 &  0.007 &  0.009 &  0.010 &  0.012 &  0.012 &  0.018 &  0.007 &  0.008 \\  
B11	&  0.002 &  0.005 &  0.006 &  0.006 &  0.008 &  0.008 &  0.010 &  0.010 &  0.015 &  0.005 &  0.007 \\  
\hline 
 \hline 
\end{tabular}
%\end{table*}

\vspace{0.4in}

%\begin{table*}[h]
\caption{The correlation matrix for the point-to-point uncertainties in the central-central dijet measurements (Topology C).  The relative luminosity and beam polarization uncertainties, which are common to all the data points, are not included.}
\label{tab:corrCVsC}
\begin{tabular}{c@{~~~}c@{~~}c@{~~}c@{~~}c@{~~}c@{~~}c@{~~}c@{~~}c@{~~}c@{~~}c}
\hline 
 \hline 
Bin	& C1 & C2 & C3 & C4 & C5 & C6 & C7 & C8 & C9 & C10 \\ 
\hline 
C1	&  1 &  0.004 &  0.003 &  0.004 &  0.004 &  0.003 &  0.003 &  0.003 &  0.002 &  0.006 \\  
C2	&   &  1 &  0.016 &  0.018 &  0.017 &  0.012 &  0.016 &  0.013 &  0.011 &  0.024 \\  
C3	&   &   &  1 &  0.019 &  0.017 &  0.012 &  0.013 &  0.010 &  0.008 &  0.017 \\  
C4	&   &   &   &  1 &  0.019 &  0.013 &  0.016 &  0.013 &  0.010 &  0.022 \\  
C5	&   &   &   &   &  1 &  0.012 &  0.015 &  0.013 &  0.010 &  0.023 \\  
C6	&   &   &   &   &   &  1 &  0.012 &  0.010 &  0.008 &  0.018 \\  
C7	&   &   &   &   &   &   &  1 &  0.015 &  0.012 &  0.029 \\  
C8	&   &   &   &   &   &   &   &  1 &  0.011 &  0.027 \\  
C9	&   &   &   &   &   &   &   &   &  1 &  0.023 \\  
C10	&   &   &   &   &   &   &   &   &   &  1 \\  
\hline 
 \hline 
\end{tabular}
\end{table*}

\begin{table*}[h]
\caption{The correlation matrix for the point-to-point uncertainties coupling the central-central dijet measurements (Topology C) with the forward-backward dijet measurements (Topology D).  The relative luminosity and beam polarization uncertainties, which are common to all the data points, are not included.}
\label{tab:corrCVsD}
\begin{tabular}{c@{~~~}c@{~~}c@{~~}c@{~~}c@{~~}c@{~~}c@{~~}c@{~~}c@{~~}c@{~~}c@{~~}c}
\hline 
 \hline 
Bin	& D1 & D2 & D3 & D4 & D5 & D6 & D7 & D8 & D9 & D10 & D11 \\ 
\hline 
C1	&  0.000 &  0.001 &  0.001 &  0.001 &  0.002 &  0.002 &  0.002 &  0.002 &  0.003 &  0.001 &  0.001 \\  
C2	&  0.001 &  0.004 &  0.005 &  0.005 &  0.006 &  0.007 &  0.008 &  0.009 &  0.013 &  0.005 &  0.006 \\  
C3	&  0.001 &  0.003 &  0.004 &  0.004 &  0.004 &  0.005 &  0.006 &  0.006 &  0.009 &  0.003 &  0.004 \\  
C4	&  0.001 &  0.004 &  0.005 &  0.005 &  0.006 &  0.007 &  0.008 &  0.008 &  0.011 &  0.004 &  0.005 \\  
C5	&  0.001 &  0.004 &  0.005 &  0.005 &  0.006 &  0.007 &  0.008 &  0.008 &  0.012 &  0.004 &  0.005 \\  
C6	&  0.001 &  0.003 &  0.004 &  0.004 &  0.005 &  0.006 &  0.006 &  0.007 &  0.010 &  0.004 &  0.004 \\  
C7	&  0.002 &  0.006 &  0.007 &  0.006 &  0.008 &  0.009 &  0.010 &  0.011 &  0.016 &  0.006 &  0.007 \\  
C8	&  0.002 &  0.005 &  0.006 &  0.006 &  0.008 &  0.008 &  0.010 &  0.010 &  0.015 &  0.005 &  0.007 \\  
C9	&  0.001 &  0.005 &  0.006 &  0.005 &  0.007 &  0.007 &  0.008 &  0.009 &  0.013 &  0.005 &  0.006 \\  
C10	&  0.003 &  0.011 &  0.014 &  0.013 &  0.016 &  0.018 &  0.021 &  0.022 &  0.032 &  0.012 &  0.015 \\  
\hline 
 \hline 
\end{tabular}
\end{table*}

\begin{table*}[h]
\caption{The correlation matrix for the point-to-point uncertainties in the forward-backward dijet measurements (Topology D).  The relative luminosity and beam polarization uncertainties, which are common to all the data points, are not included.}
\label{tab:corrDVsD}
\begin{tabular}{c@{~~~}c@{~~}c@{~~}c@{~~}c@{~~}c@{~~}c@{~~}c@{~~}c@{~~}c@{~~}c@{~~}c}
\hline 
 \hline 
Bin	& D1 & D2 & D3 & D4 & D5 & D6 & D7 & D8 & D9 & D10 & D11 \\ 
\hline 
D1	&  1 &  0.004 &  0.012 &  0.007 &  0.007 &  0.005 &  0.005 &  0.003 &  0.003 &  0.001 &  0.001 \\  
D2	&   &  1 &  0.009 &  0.006 &  0.007 &  0.006 &  0.006 &  0.005 &  0.007 &  0.003 &  0.003 \\  
D3	&   &   &  1 &  0.014 &  0.015 &  0.012 &  0.011 &  0.008 &  0.010 &  0.004 &  0.004 \\  
D4	&   &   &   &  1 &  0.010 &  0.008 &  0.008 &  0.007 &  0.008 &  0.003 &  0.004 \\  
D5	&   &   &   &   &  1 &  0.010 &  0.010 &  0.008 &  0.010 &  0.004 &  0.004 \\  
D6	&   &   &   &   &   &  1 &  0.009 &  0.008 &  0.011 &  0.004 &  0.005 \\  
D7	&   &   &   &   &   &   &  1 &  0.009 &  0.012 &  0.005 &  0.006 \\  
D8	&   &   &   &   &   &   &   &  1 &  0.012 &  0.005 &  0.006 \\  
D9	&   &   &   &   &   &   &   &   &  1 &  0.007 &  0.008 \\  
D10	&   &   &   &   &   &   &   &   &   &  1 &  0.003 \\  
D11	&   &   &   &   &   &   &   &   &   &   &  1 \\  
\hline 
 \hline 
\end{tabular}
\end{table*}

%\vspace{0.2in}
\begin{acknowledgments}
We thank the RHIC Operations Group and RCF at BNL, the NERSC Center at LBNL, and the Open Science Grid consortium for providing resources and support.  This work was supported in part by the Office of Nuclear Physics within the U.S. DOE Office of Science, the U.S. National Science Foundation, the Ministry of Education and Science of the Russian Federation, National Natural Science Foundation of China, Chinese Academy of Science, the Ministry of Science and Technology of China and the Chinese Ministry of Education, the National Research Foundation of Korea, Czech Science Foundation and Ministry of Education, Youth and Sports of the Czech Republic, Hungarian National Research, Development and Innovation Office (FK-123824), New National Excellency Programme of the Hungarian Ministry of Human Capacities (UNKP-18-4), Department of Atomic Energy and Department of Science and Technology of the Government of India, the National Science Centre of Poland, the Ministry  of Science, Education and Sports of the Republic of Croatia, RosAtom of Russia and German Bundesministerium fur Bildung, Wissenschaft, Forschung and Technologie (BMBF) and the Helmholtz Association.
\end{acknowledgments}

\clearpage

\bibliography{Guide2Authors}
\end{document}